\documentclass[11pt]{article}
\usepackage{graphicx,color}  
\usepackage{dcolumn}   
\usepackage{bm}        
\usepackage{amssymb} 
\usepackage{amsmath}
\usepackage{braket}
\usepackage{cancel}
\usepackage{subfigure}
\usepackage{fullpage}

 \newcommand{\beq}{\begin{equation}}
\newcommand{\eeq}{\end{equation}}
\newcommand{\beqa}{\begin{eqnarray}}
\newcommand{\eeqa}{\end{eqnarray}}
\newcommand{\be}{\begin{equation}}
\newcommand{\ee}{\end{equation}}

\newcommand{\appref}[1]{Appendix~\ref{#1}}		

\newcommand{\nc}{\newcommand}
\nc{\rnc}{\renewcommand}
\rnc{\d}{\mathrm{d}}
\nc{\D}{\partial}
\nc{\K}{\kappa}
\nc{\bK}{\bar{\K}}
\nc{\bN}{\bar{N}}
\nc{\bq}{\bar{q}}
\nc{\vbq}{\vec{\bar{q}}}
\nc{\g}{\gamma}
\nc{\lrarrow}{\leftrightarrow}
\nc{\rg}{\sqrt{g}}
\nc{\nn}{\nonumber}
\rnc{\(}{\left(}
\rnc{\)}{\right)}
\nc{\q}{\vec{q}}
\nc{\x}{\vec{x}}
\rnc{\a}{\hat{a}}
\nc{\ep}{\epsilon}
\nc{\tto}{\rightarrow}
\rnc{\inf}{\infty}
\rnc{\Re}{\mathrm{Re}}
\rnc{\Im}{\mathrm{Im}}
\nc{\z}{\zeta}
\nc{\mA}{\mathcal{A}}
\nc{\mB}{\mathcal{B}}
\nc{\mC}{\mathcal{C}}
\nc{\mD}{\mathcal{D}}
\nc{\mN}{\mathcal{N}}
\rnc{\H}{\mathcal{H}}
\rnc{\L}{\mathcal{L}}
\nc{\<}{\langle}
\rnc{\>}{\rangle}
\nc{\fnl}{f_{NL}}
\nc{\gnl}{g_{NL}}
\nc{\fnleq}{f_{NL}^{equil.}}
\nc{\fnlloc}{f_{NL}^{local}}
\nc{\vphi}{\varphi}
\nc{\Lie}{\pounds}
\nc{\half}{\frac{1}{2}}
\nc{\bOmega}{\bar{\Omega}}
\nc{\bLambda}{\bar{\Lambda}}
\nc{\dN}{\delta N}
\nc{\gYM}{g_{\mathrm{YM}}}
\nc{\geff}{g_{\mathrm{eff}}}
\nc{\tr}{\mathrm{tr}}
\nc{\oa}{\stackrel{\leftrightarrow}}
\nc{\IR}{{\rm IR}}
\nc{\UV}{{\rm UV}}
\nc{\la}{\lambda}
\nc{\veps}{\varepsilon}
\begin{document}
	\begin{center}
		\vspace{1.5cm}
		{\Large \bf Parity-Odd 3-Point Functions from CFT  in Momentum Space\\ and the Chiral Anomaly
		 \\ }
		\vspace{0.3cm}
		
		\vspace{1cm}
		{\bf Claudio Corian\`o$^{(1)}$, Stefano Lionetti$^{(1)}$ and Matteo Maria Maglio$^{(2)}$ \\}
		\vspace{1cm}
		{\it  $^{(1)}$Dipartimento di Matematica e Fisica, Universit\`{a} del Salento \\
			and INFN Sezione di Lecce, Via Arnesano 73100 Lecce, Italy\\
			National Center for HPC, Big Data and Quantum Computing\\}
		\vspace{0.5cm}
		{\it  $^{(2)}$ Institute for Theoretical Physics (ITP), University of Heidelberg\\
			Philosophenweg 16, 69120 Heidelberg, Germany}

		\begin{abstract}
			We illustrate how the Conformal Ward Identities (CWI) in momentum space for parity-odd correlators determine the structure of a chiral anomaly interaction, taking the example of the VVA (vector/vector/axial-vector) and AAA correlators in momentum space. Only the conservation and the anomalous WIs, together with the Bose symmetry, are imposed from the outset for the determination of the correlators. We use a longitudinal/transverse decomposition of tensor structures and form factors.  The longitudinal (L) component is fixed by the anomaly content and the anomaly pole, while in the transverse (T) sector we define a new parameterization. We relate the latter both to the Rosenberg original representation of the VVA and to the longitudinal/transverse (L/T) one, first introduced in the analysis of $g-2$ of the muon in the investigation of the diagram in the chiral limit of QCD.  The  correlators are completely identified by the conformal constraints whose solutions are fixed only by the anomaly coefficient, the residue of the anomaly pole. In both cases, our CFT result matches the one-loop perturbative expression, as expected. The CWIs for correlators of mixed chirality $J_L J J_R$ generate solutions in agreement with the all-orders nonrenormalization theorems of perturbative QCD and in the chiral limit. 
			  				
\end{abstract}
	\end{center}
	\titlepage
	
	\section{Introduction}
Parity even correlators in $d=4$ play a central role in CFT and have been investigated in coordinate space \cite{Osborn:1993cr,Erdmenger:1996yc} for quite some time. Parity odd ones, instead, have attracted less attention, except for the $JJJ_5$ and $J_5J_5J_5$ (or $VVA$ and $AAA$, where $A$ and $V$ refer to the axial-vector and to the vector current respectively) due to the role of the chiral anomaly. \\
 The conformal properties of the VVA diagram, in the coordinate space, have been studied since the 70's (see for instance \cite{Schreier:1971um} and \cite{Erlich:1996mq}). 
In more recent years, its non-renormalization property, which affects only its anomaly (longitudinal) part, as shown by Adler and Bardeen \cite{Adler:1969er}, has been redrawn in a more general context, given its relevance in the analysis of $g-2$ of the muon \cite{Vainshtein:2002nv}. The numerical value of $a_\mu=(g-2)_\mu=F_\mathrm{M}(0)$, which measures the muon anomaly, is given by the Pauli form factor at zero
momentum transfer, and involves a soft photon limit on one of the two vector currents of the $VVA$ diagram, which interpolates with $F_\mathrm{M}$ via electroweak corrections.\\
In such special kinematic configuration, the correlator  is described by the longitudinal and transverse components $w_L$ and $w_T$ of the diagram.
  If we adopt such a $L/T$  separation of the vertex, $w_L$ does not receive radiative corrections, as shown in \cite{Adler:1969er}. Therefore, any constraining relation involving both $w_L$ and $w_T$ is bound to identify combinations of form factors in the $T$ sector, which are protected against radiative corrections. The conditions under which such nonrenormalization constraints hold, may involve a special kinematics. For this reason, the analysis of the correlator requires moving into momentum space. \\
 It has been argued that in the chiral limit of QCD and with a soft $V$ line, the relation  
   \beq
 \label{lo}
 w_L={2} \, w_T
 \eeq
 holds to all orders in the strong coupling constant $\alpha_s$ 
 \cite{Vainshtein:2002nv}. 
 This constraint is indeed reproduced by the bare fermion loop, which is conformal,  and is a byproduct of our analysis of the CWIs in momentum space as well.  
 Notice that, given the complete overlap of the CFT result with the perturbative one at the lowest order, identities such as the Crewther-Broadhurst-Kataev (CBK) relation \cite{Crewther:1972kn,Broadhurst:1993ru, Gabadadze:1995ei}, are exactly satisfied in our case as well (see \cite{Gabadadze:2017ujx}). The CBK identity connects the Adler function for electron-positron annihilation with the Bjorken function of deep-inelastic sum rules. Away from the conformal limit, these relations acquire corrections proportional to the QCD $\beta$ function at higher orders.\\
A perturbative analysis of the radiative corrections of the $VVA$ correlator in QCD at two-loop level ($O(\alpha_s$),  showed that the entire vertex is also non renormalized \cite{Jegerlehner:2005fs}, in agreement with \eqref{lo}. However, at higher orders in $\alpha_s$ ($O(\alpha_s^2)$), the nonrenormalization properties of the entire diagram are violated and conformality is lost. In particular 
 \eqref{lo} is only limited to the soft photon limit discussed in \cite{Vainshtein:2002nv}, as shown in explicit computations \cite{Mondejar:2012sz}. \\
In our analysis, the constraint in \eqref{lo} can be viewed also as a result of conformal symmetry, without resorting to a perturbative picture, since in the unique solution of the CWIs from momentum space, such constraint is verified.
While the result is expected, the procedure is novel.  \\
One interesting aspect of the identification of the $VVA$ or $AAA$ diagrams from the CWIs in momentum space, 
is the centrality of the anomaly pole in the longitudinal sector, that allows to identify every part of the correlator. The link between such massless state and the nonlocal anomaly effective action has been stressed in the past in several works, \cite{Giannotti:2008cv} \cite{2009PhLB..682..322A} \cite{Coriano:2014gja}. More recently, these analysis have been extended in the investigation of chiral density waves, with local actions \cite{Mottola:2019nui}\cite{Ferrer:2020ulz}. These actions play a key role in very different contexts where either chiral or conformal anomalies are involved, such as in topological materials \cite{Chernodub:2021nff} 
\cite{Arjona:2019lxz}.

\subsection{Conformal analysis in momentum space }
 
In coordinate space the identification of the structure of conformal correlators is quite direct, but does not provide much information on the dynamical properties of those, among them, which are affected by an anomaly. \\
In momentum space the vertices of such interactions have been investigated, in perturbation theory, in terms of different (minimal and non minimal) sets of form factors, whose expressions can be calculated at one-loop level by standard perturbative methods. At $d=4$ the Schouten identities play an important role in the choice of the most 
useful representations of such diagrams, and relate the form factors of the different parameterizations.
For instance, the $w_L/w_T$ decomposition, that plays a fundamental role in the investigations mentioned above, can be related to other representations of the corresponding form factors.\\
Being a free fermion Lagrangian at $d=4$ conformal at tree level, the chiral anomaly fermion loop provides a simple free field theory realization of the VVA vertex, which is also conformal. As we have mentioned, the conformal symmetry is eventually broken by radiative corrections only at order $\alpha_s^2$, in the strong coupling constant. 
Anomalies arise from the region of the correlator where all the external coordinate points $(x,y,z)$ coalesce, and for this reason, by a Fourier transform, their analysis in momentum space has the advantage of including this region rather automatically. While the conformal contributions to 
such correlators at $d=4$ are described, as mentioned, by the simple massless fermion loop, the identification of the kinematical structure of the vertex in momentum space - without any reference to a Feynman diagram realization and using only the Ward identities (WIs) of the case - has never been discussed before. This is the goal of our work.\\ We are going to close this gap and show how the entire correlator is fixed by the anomaly also if we resort to a momentum space analysis. We will be using a variant of the separation of the CWIs into primary and secondary equations \cite{Bzowski:2013sza}, in which the anomaly constraint on the longitudinal form factor is imposed from the beginning. This approach differs from the cases discussed so far for 
3-point and 4-point functions affected by the conformal anomaly, since in that case the anomaly emerges from the renormalization procedure, realized by the inclusion of a Weyl squared counterterm after renormalization (see \cite{Coriano:2018bbe,Coriano:2018bsy} for a Lagrangian formulation). 
Renormalized parity even correlators, type-A and type-B Weyl anomalies have been discussed in \cite{Bzowski:2017poo,Bzowski:2018fql}.
It is well known that for a VVA vertex, as for any anomaly interaction which is purely topological, the process of renormalization can be entirely bypassed by imposing the anomalous WIs on the external currents. 
Our approach extends to chirally odd three-point functions the methods used in the analysis of the parity even ones in momentum space \cite{Bzowski:2013sza,Coriano:2013jba}, in the presence of a chiral anomaly, and can be formulated also for higher point functions. \\
Recently novel approaches have been adopted for the construction of parity odd correlation functions. In particular in \cite{Jain:2021gwa} it is shown that parity odd CFT 3-point functions can be obtained by doing epsilon transformation starting from parity even CFT correlation functions.
Moreover, in \cite{Jain:2021wyn} the authors use both the momentum space CWIs as well as spin-raising and weight-shifting operators to fix the form of parity odd correlators. However both \cite{Jain:2021gwa} and \cite{Jain:2021wyn} do not consider anomalous correlators. \\
A recent analysis of 3-point functions for parity even correlators with non-conserved currents of arbitrary spin has been discussed in \cite{Marotta:2022jrp}. Moreover, another recent analysis on the conformal bootstrap equations in momentum space at finite volume has been discussed in \cite{Nishikawa:2023zcv}.

\subsection{Massless intermediate states in the anomaly: the pivot}
The construction of the entire correlator proceeds from the anomaly pole, that plays a key role in any anomaly diagram. This takes the role of a pivot in the procedure, and it allows to solve for the longitudinal anomalous WI quite straightforwardly. \\
The tensorial expansion of a chiral vertex is not unique, due to the presence of Schouten relations among its tensor components. For instance, an anomaly pole in the virtuality of the axial-vector current - denoted as $1/p_3^2 $ in our notations -  can be inserted or removed from a given tensorial decomposition, just by the use of these relations. \\
These identities connect two of the most common representation of this vertex, the first of them introduced long ago 
by Rosenberg \cite{Rosenberg:1962pp}, expressed in terms of 6 tensor structures and form factors, and the second one \cite{Knecht:2003xy}, more recent, introduced in the context of the analysis of the $g-2$ anomalous magnetic moment of the muon. The latter, which is the most valuable from the physical point of view, allows to attribute the anomaly to the exchange of a pole in the longitudinal channel \cite{Dolgov:1971ri} \cite{Giannotti:2008cv,Armillis:2009pq}. \\
In this second parameterization of the vertex, the decomposition identifies longitudinal and transverse components 
\beq
\langle J^{\mu_1}(p_1)J^{\mu_2}(p_2)J^{\mu_3}_5(p_3) \rangle=\frac{1}{8\pi^2}\left(
W_L^{\mu_1\mu_2\mu_3}- W_T^{\mu_1\mu_2\mu_3}\right)
\label{sec}
\eeq
where $W_T$ is the transverse part, while the longitudinal tensor structure is given by
\beq
\label{refe}
W_L^{\mu_1\mu_2\mu_3}=w_L \, {p_3^{\mu_3}}\epsilon^{\mu_1\mu_2 \rho\sigma}{p_{1\rho}p_{2\sigma}}\equiv w_L\,
{p_3^{\mu_3}}\epsilon^{\mu_1\mu_2  p_1 p_2}
\eeq
$w_L$ is the anomaly form factor, that 
in the massless (chiral or conformal) has a ${1}/{p_3^2}$ pole.
In the case of gauge anomalies, the cancelation of the anomaly poles is entirely connected with the particle content of the theory and defines the condition for the elimination of such massless interactions. The total residue at the pole then identifies the total anomaly of a certain fermion multiplet.\\
 A similar behaviour holds for conformal correlators with stress energy tensors, where the residue at the pole coincides with the $\beta$-function of the Lagrangian field theory, and is determined by the number of massless degrees of freedom included in the corresponding anomaly vertex, at the scale at which the perturbative prescription holds \cite{Coriano:2014gja}.\\
 \begin{figure}[t]
\centering
\subfigure[]{\includegraphics[scale=0.8]{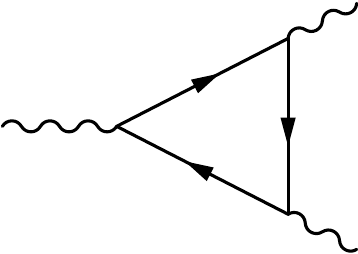}}  \hspace{2cm}
\subfigure[]{\includegraphics[scale=0.8]{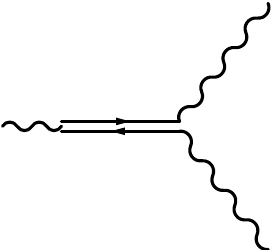}} \hspace{2cm}
\subfigure[]{\includegraphics[scale=0.8]{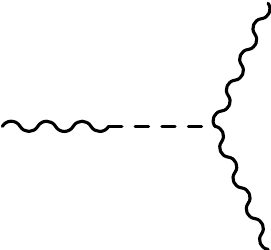}}
\caption{The fermion loop (a); the collinear region (b); the effective pseudoscalar exchange (c).}
\label{ddiag}
\end{figure}
 As illustrated in Fig. 1,  the pole emerges from the region of integration in momentum space in the VVA loop when an effective interaction, mediated by the fermion/antifermion pair is generated. The two collinear (on-shell) particles describe an effective pseudoscalar interaction interpolating between the incoming axial vector and the two outgoing vector currents. \\
Away from the conformal limit, when the VVA diagram is recomputed with the inclusion of a fermion of mass $m$,
one discovers the presence of a sum rule satisfied by the spectral density of the form factor $w_L$ of the longitudinal channel. Such form factor $(w_L)$ is characterised by a spectral density $\rho(s)$, whose integration in the region $4 m^2 < s < \infty$ in the dispersive variable $s$, related to the virtuality of the axial-vector current, is 
mass-independent and given by 
\beq
\int_{4 m^2}^\infty d s\, \rho(s,m^2)=a
\label{deltaf}
\eeq   
where $a$ is the anomaly for a single Dirac fermion. For on-shell vector lines the perturbative vertex reduces only to the longitudinal component $W_L$.\\
The separation into transverse and longitudinal contributions appears to be ambiguous, in the sense that one could always add and subtract transverse contributions to the pole part. However, one can show that such $1/p_3^2$ behaviour does not depend on the parameterization of the fermion loop. This important point is illustrated in the case of the longitudinal transverse (L/T) representation, once this representation is mapped to the Rosenberg representation, as we are going to comment below.
A perturbative reparameterization of the loop momentum in the anomaly diagram is equivalent to the inclusion of Chern-Simons terms in the representation of two of the six form factors ($B_1$ and $B_2$) that we are going to define in the next sections, but it is independent of 
them in the second parameterization ($w_L\sim B_1- B_2$). \\
Our analysis shows the centrality of the pole in determining the general solution of the CWIs, that is only fixed by the anomaly. Some of these points are reviewed in the appendix for completeness.
The procedure follows a decomposition similar to the one used in the case of correlators of stress energy tensors. It is based on the parameterization of such correlators in terms of a transverse traceless and a longitudinal sector. In the conformal case the trace anomaly sector of the correlator, identified by pole structures similar to the one that we are going to discuss here, needs to be modified by the inclusion of traceless Weyl invariant terms \cite{Coriano:2021nvn,Coriano:2022jkn}. These are necessary to decompose the correlator into two traceless and trace parts which both satisfy energy momentum conservation. In the case of the chiral VVA the procedure is far simpler. \\
In the case of the AAA diagram, as we are going to see, we are going to encounter contributions which are proportional to Chern-Simons (CS) forms.  These are terms linear in two of the three momenta and allow to move the anomaly form one vertex to the other. For instance, in a perturbative evaluation of the $VVA$ and $AAA$ diagrams, in the chiral limit, the anomaly can be moved around the vertices by the inclusion of such CS terms 
\beq \label{eq:VVAAAACS}
\langle J^{\mu_1}J^{\mu_2}J_5^{\mu_3}\rangle =\langle J_5^{\mu_1}J_5^{\mu_2}J_5^{\mu_3}\rangle+\frac{8 \, a \, i }{3}\varepsilon^{\alpha \mu_1\mu_2\mu_3}(p_{1}-p_{2})_\alpha.
 \eeq
The term $\varepsilon^{\alpha \mu_1\mu_2\mu_3}(p_{1}-p_{2})_\alpha$ corresponds to the Chern-Simons contribution. \\
In our analysis we proceed from the $VVA$ case, then turning to the $AAA$ and show in both cases how primary and secondary CWI can be solved quite efficiently 
in momentum space. As shown in Fig. 2, the structure of the $AAA$ is dictated by symmetry. One can move around the anomaly content, from the $AAA$ with CS terms in order to obtain the VVA  \cite{Armillis:2007tb}.
 Our solutions are then compared with the two most common representations of such diagrams, the Rosenberg and the L/T one. The latter is particularly useful for the role that provides for the longitudinal component of the correlator. The same representation has been used in the past for the derivation of new nonrenormalization theorems in perturbative QCD, in the chiral limit. This result has been derived in \cite{Knecht:2003xy} and is confirmed by the formal analysis of the CWIs, once we turn to discuss the relation between an $AVV$ and a $VVA$ diagram, whose difference, in our analysis, is found to be expressed in terms of a CS form. For convenience, we have included an appendix where some of the more technical aspects of our derivation are collected. 

\begin{figure}[t]
{\centering \resizebox*{12cm}{!}{\rotatebox{0}
{\includegraphics{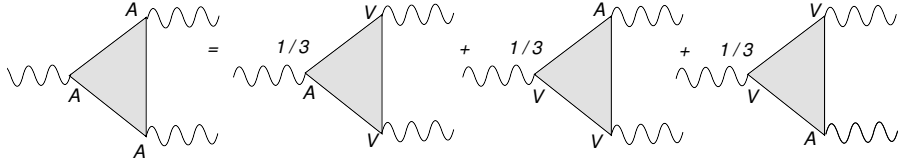}}}\par}
\caption{Distribution of the axial anomaly for the ${\bf AAA}$ diagram}
\label{distribution}
\end{figure}

\section{The conformal $\langle VVA \rangle$ correlator}
\subsection{Longitudinal/Transverse decomposition}
The analysis of the conformal constraints for $\braket{JJJ_5}$, as already mentioned, is performed by applying the L/T decomposition to the 
correlator. 
 We focus our analysis on the $d=4$ case, where the conformal dimensions of the conserved currents $J^\mu$ are $\Delta=3$ and the 
 tensorial structures of the correlator will involve the antisymmetric tensor in four dimensions $\epsilon^{\mu\nu\alpha\beta}$.  The procedure to obtain the general structure of the correlator starts from the conservation Ward identities
\begin{align}\label{eq:CWIJJJ}
\nabla_\mu \braket{J^\mu}=0, \qquad\qquad \nabla_\mu \braket{J^\mu_5}=a \, \varepsilon^{\mu\nu\rho\sigma}F_{\mu\nu}F_{\rho\sigma}
\end{align}
of the expectation value of the non anomalous $J^{\mu}$ and anomalous $J_5^{\mu}$ currents. The vector currents are coupled to the 
vector source $V_{\mu}$ and the axial-vector current to the source $A_\mu$.
Applying multiple functional derivatives to \eqref{eq:CWIJJJ} with respect to the source $V_{\mu}$, after a Fourier transform, we find the conservation Ward identities related to the entire correlator which are given by
\begin{equation}\label{eq:jjjconsward3p}
\begin{aligned}
&p_{i\mu_i}\,\braket{J^{\mu_1}(p_1)J^{\mu_2}(p_2)J_5^{\mu_3}(p_3)}=0,\quad \quad i=1,2\\[1.2ex]
&p_{3\mu_3}\,\braket{J^{\mu_1}(p_1)J^{\mu_2}(p_2)J_5^{\mu_3}(p_3)}=-8 \, a \, i \, \varepsilon^{p_1p_2\mu_1\mu_2}
\end{aligned}
\end{equation}
where $\varepsilon^{p_1p_2\mu_1\mu_2}\equiv\varepsilon^{\alpha\beta\mu_1\mu_2}p_{1\alpha}p_{2\beta}$ and the momenta are all incoming.
From this relations we construct the general form of the correlator, splitting the operators into 
a transverse and a longitudinal part as
\begin{equation}
\begin{aligned}
\label{ex}
&J^{\mu}(p)=j^\mu(p)+j_{loc}^\mu(p),\\
&j^{\mu}=\pi^{\mu}_{\alpha}(p)\,J^{\alpha}(p),\quad \pi^{\mu}_\alpha(p)\equiv \delta^{\mu}_\alpha-\frac{p_\alpha\,p^\mu}{p^2},\\
&j_{loc}^{\mu}(p)=\frac{p^\mu}{p^2}\,p\cdot J(p)
\end{aligned}
\end{equation}
and
\begin{equation}
\begin{aligned}
\label{ex}
&J_5^{\mu}(p)=j_5^\mu(p)+j_{5 loc}^\mu(p),\\
&j_5^{\mu}=\pi^{\mu}_{\alpha}(p)\,J_5^{\alpha}(p),\\
&j_{ 5 loc}^{\mu}(p)=\frac{p^\mu}{p^2}\,p\cdot J_5(p)
\end{aligned}
\end{equation}
Due to \eqref{eq:jjjconsward3p}, the correlator is purely transverse in the vector currents. We then have the following decomposition
\begin{equation}
	\braket{ J^{\mu_1 }(p_1) J^{\mu_2 } (p_2)J_5^{\mu_3}(p_3)}=\braket{ j^{\mu_1 }(p_1) j^{\mu_2 }(p_2) j_5^{\mu_3}(p_3)}+\braket{J^{\mu_1 }(p_1) J^{\mu_2 }(p_2)\, j_{5 \text { loc }}^{\mu_3}(p_3)}\label{decomp}
\end{equation}
where the first term is completely transverse with respect to the momenta $p_{i\mu_i}$, $i=1,2,3$ and the second term is the longitudinal part that is proper of the anomaly contribution. Using the anomaly constraint on $j_{5 loc}$ we obtain
\begin{align}
\braket{J^{\mu_1 }(p_1) J^{\mu_2 }(p_2)\, j_{5 \text { loc }}^{\mu_3}(p_3)}=\frac{p_3^{\mu_3}}{p_3^2}\,p_{3\,\alpha_3}\,\braket{J^{\mu_1}(p_1)J^{\mu_2}(p_2)J_5^{\alpha_3}(p_3)}=-\frac{8\,a\,i}{p_3^2}\varepsilon^{p_1p_2\mu_1\mu_2}\,p_3^{\mu_3}
\end{align}
On the other hand, the transverse part can be formally written as
\begin{align}
	\braket{j^{\mu_1}(p_1)j^{\mu_2}(p_2) j^{\mu_3}_5 (p_3)} &=\pi^{\mu_1}_{\alpha_1}(p_1)
	\pi^{\mu_2}_{\alpha_2} (p_2) \pi^{\mu_3}_{\alpha_3}
	\left(p_3\right)\Big[ 	A_1(p_1,p_2,p_3)\,\varepsilon^{p_1p_2\alpha_1\alpha_2}p_1^{\alpha_3} +A_2(p_1,p_2,p_3)\,\varepsilon^{p_1p_2\alpha_1\alpha_3}p_3^{\alpha_2} \notag\\
	& +A_3(p_1,p_2,p_3)\,\varepsilon^{p_1p_2\alpha_2\alpha_3}p_2^{\alpha_1} + 
	A_4(p_1,p_2,p_3)\, \varepsilon^{p_1 \alpha_1\alpha_2\alpha_3}  +
	A_5(p_1,p_2,p_3)\, \varepsilon^{p_2\alpha_1\alpha_2\alpha_3}  
	\Big]
	\label{decomp1}
	\end{align}
where we have made a choice on which independent momenta to consider for each index, and in particular
\begin{align}
\alpha_1\leftrightarrow\,p_1,p_2, \qquad \alpha_2\leftrightarrow\,p_2,p_3,\qquad \alpha_3\leftrightarrow\,p_3,p_1
\end{align}
The correlator has to be symmetric under the exchange of the two vector currents and this fact is reflected in the symmetry constraints
\begin{align}
A_3(p_1,p_2,p_3)=-A_2(p_1,p_2,p_3),\quad A_5(p_1,p_2,p_3)=-A_4(p_1,p_2,p_3),\quad A_1(p_1,p_2,p_3)=-A_1(p_1,p_2,p_3)
\end{align}
reducing by two the number of independent form factors. Furthermore, in $d=4$ a class of tensor identities has to be considered, for instance the Schouten identity
\begin{align}
\delta^{\beta_3 [\alpha_1} \varepsilon^{\alpha_2 \alpha_3 \beta_1 \beta_2]}=0.
\end{align}
From this type of tensor identities we find that
\begin{align}
\pi^{\mu_1}_{\alpha_1}
\pi^{\mu_2}_{\alpha_2} \pi^{\mu_3}_{\alpha_3}\bigg(p_2^{\alpha_1} \varepsilon^{p_1 p_2 {\alpha_2} {\alpha_3}}\bigg)&=\pi^{\mu_1}_{\alpha_1}
\pi^{\mu_2}_{\alpha_2} \pi^{\mu_3}_{\alpha_3}\bigg(-\big(p_2\cdot p_1\big) \varepsilon^{p_2\alpha_1 \alpha_2\alpha_3}+p_2^2\,\varepsilon^{p_1\alpha_1\alpha_2\alpha_3}-p_2^{\alpha_3}\,\varepsilon^{ p_1p_2\alpha_1\alpha_2}\bigg),\\
\pi^{\mu_1}_{\alpha_1}
\pi^{\mu_2}_{\alpha_2} \pi^{\mu_3}_{\alpha_3}\bigg(p_3^{\alpha_2}\varepsilon^{p_1 p_2 \alpha_1\alpha_3}\bigg)&=\pi^{\mu_1}_{\alpha_1}
\pi^{\mu_2}_{\alpha_2} \pi^{\mu_3}_{\alpha_3}\bigg(-p_1^2\,\varepsilon^{p_2 \alpha_1\alpha_2\alpha_3}+\big(p_1\cdot p_2\big)\varepsilon^{p_1\alpha_1\alpha_2\alpha_3}-p_1^{\alpha_3}\varepsilon^{p_1p_2\alpha_1\alpha_2}\bigg),
\end{align}
reducing the number of independent form factors just to two. We conclude that the general structure of the transverse part is given by
\begin{align}
	\braket{j^{\mu_1}(p_1)j^{\mu_2}(p_2) j^{\mu_3}_5 (p_3)} &=\pi^{\mu_1}_{\alpha_1}(p_1)
	\pi^{\mu_2}_{\alpha_2} (p_2) \pi^{\mu_3}_{\alpha_3}
	\left(p_3\right)\Big[ 	A_1(p_1,p_2,p_3)\,\varepsilon^{p_1p_2\alpha_1\alpha_2}p_1^{\alpha_3} \notag\\
	& \qquad+ 
	A_2(p_1,p_2,p_3)\, \varepsilon^{p_1 \alpha_1\alpha_2\alpha_3}  -
	A_2(p_2,p_1,p_3)\, \varepsilon^{p_2\alpha_1\alpha_2\alpha_3}  
	\Big]\label{decompFin}
\end{align}
where $	A_1(p_1,p_2,p_3)=-A_1(p_2,p_1,p_3)$. 

\subsection{Dilatation Ward identities}

We now start to analyse the conformal constraints on the form factors. In this and the next sections, we closely follows the methodology adopted in \cite{Bzowski:2013sza}.
The invariance of the correlator under dilatation is reflected in the equation
\begin{align}
\left(\sum_{i=1}^3\Delta_i-2d-\sum_{i=1}^2\,p_i^\mu\frac{\partial}{\partial p_i^\mu}\right)\braket{J^{\mu_1}(p_1)J^{\mu_2}(p_2) J^{\mu_3}_5 (p_3)}=0.\label{Diltt}
\end{align}
Considering the decomposition \eqref{decomp} in the previous equation, then the transverse part of the correlator has to satisfy 
\begin{align}
	\left(\sum_{i=1}^3\Delta_i-2d-\sum_{i=1}^2\,p_i^\mu\frac{\partial}{\partial p_i^\mu}\right)\braket{j^{\mu_1}(p_1)j^{\mu_2}(p_2) j^{\mu_3}_5 (p_3)}=0.
\end{align}
By using the chain rule
\begin{align}
\frac{\partial}{\partial p_i^\mu}=\sum_{j=1}^3\frac{\partial p_j}{\partial p_i^\mu}\frac{\partial}{\partial p_j}
\end{align}
in term of the invariants $p_i=|\sqrt{p_i^2}|$, we rewrite \eqref{Diltt},considering $\Delta_1=\Delta_2=d-1$, for the form factors in order to obtain
		\begin{align}
		\sum_{i=1}^{3} p_i \frac{\partial A_j}{\partial p_i }-\left(\Delta_3-2-N_j\right) A_j=0,
	\end{align}
with $N_j$ the number of momenta that the form factors multiply in the decomposition \eqref{decompFin} and then we have $N_1=3$ and $N_2=1$.

\subsection{Special Conformal Ward identities}

The invariance of the correlator with respect to the special conformal transformations is encoded in the special conformal Ward identities 
\begin{align}
0=&\sum_{j=1}^2\left[-2\frac{\partial}{\partial p_{j\kappa}}-2p_j^\alpha\frac{\partial^2}{\partial p_j^\alpha\,\partial p_{j\kappa}}+p_j^\kappa\frac{\partial^2}{\partial p_j^\alpha\,\partial p_{j\alpha}}\right]\braket{J^{\mu_1}(p_1)J^{\mu_2}(p_2)J_5^{\mu_3}(p_3)}\notag\\
&+2\left(\delta^{\mu_1\kappa}\,\frac{\partial}{\partial p_1^{\alpha_1}}-\delta^\kappa_{\alpha_1}\,\frac{\partial}{\partial p_{1\mu_1}}\right)\braket{J^{\alpha_1}(p_1)J^{\mu_2}(p_2)J_5^{\mu_3}(p_3)}\notag\\
&+2\left(\delta^{\mu_2\kappa}\,\frac{\partial}{\partial p_2^{\alpha_2}}-\delta^\kappa_{\alpha_2}\,\frac{\partial}{\partial p_{2\mu_2}}\right)\braket{J^{\mu_1}(p_1)J^{\alpha_2}(p_2)J_5^{\mu_3}(p_3)}\equiv \mathcal{K}^\kappa\braket{J^{\mu_1}(p_1)J^{\mu_2}(p_2)J_5^{\mu_3}(p_3)}.
\end{align}
The special conformal operator $\mathcal{K}^\kappa$ acts as an endomorphism on the transverse sector of the entire correlator. 
Therefore we can perform a transverse projection  ($3-\pi$ projection) on all the indices in order to identify a set of partial differential equations  
corresponding to the primary constraints \cite{Bzowski:2013sza}
\begin{align}
	&\pi_{\mu_1}^{\lambda_1}(p_1)
	\pi_{\mu_2}^{\lambda_2} (p_2) \pi_{\mu_3}^{\lambda_3}
	(p_3) \bigg(\mathcal{K}^\kappa\braket{J^{\mu_1}(p_1)J^{\mu_2}(p_2)J_5^{\mu_3}(p_3)}\bigg)=0
\end{align}
splitting the correlator into its transverse and longitudinal parts.  
Using the Schouten identities listed in \appref{appendixA}, we decompose the action of the special conformal operator on the transverse sector  using the transverse projectors. One can show that the action of the operator is endomorphic in the transverse sector. Therefore we have
\begin{equation}
\pi_{\mu_1}^{\lambda_1}(p_1)
\pi_{\mu_2}^{\lambda_2} (p_2) \pi_{\mu_3}^{\lambda_3}
(p_3) \bigg(\mathcal{K}^\kappa\braket{j^{\mu_1}(p_1)j^{\mu_2}(p_2)j_5^{\mu_3}(p_3)}\bigg)=\pi_{\mu_1}^{\lambda_1}(p_1)
\pi_{\mu_2}^{\lambda_2} (p_2) \pi_{\mu_3}^{\lambda_3}
(p_3) \,X^{\kappa\mu_1\mu_2\mu_3},\label{kjjj}
\end{equation}
where the tensor $X^{\kappa\mu_1\mu_2\mu_3}$ can be constructed using a set of possible tensor structures, of the form 
\begin{align}
\varepsilon^{ \mu_1\mu_2\mu_3p_1}p_1^\kappa,\quad
		\varepsilon^{\mu_1\mu_2\mu_3p_2}p_1^\kappa,\quad
		\varepsilon^{\mu_1\mu_2p_1p_2}p_1^{\mu_3}p_1^\kappa,\quad
		\varepsilon^{\mu_1\mu_3p_1p_2}p_3^{\mu_2}p_1^\kappa,\quad
		\varepsilon^{\mu_2\mu_3p_1p_2}p_2^{\mu_1}p_1^\kappa,
\end{align}
whose complete list is given in Appendix A. These tensor structures are not all independent, and are simplified using some Schouten identites in order to find the minimal number of tensor structures in which $X$ can be expanded. The first two identities are 
\begin{align}
\varepsilon^{[\mu_2\mu_3\kappa p_1}\delta^{\mu_1]\alpha}&=0,\\
\varepsilon^{[\mu_2\mu_3\kappa p_2}\delta^{\mu_1]\alpha}&=0,
\end{align}
that can be contracted with $p_{1\alpha}$ and $p_{2\alpha}$ to generate, after the projection, tensor identities of the form

\begin{align}
	&\pi_{\mu_1}^{\lambda_1}
	\pi_{\mu_2}^{\lambda_2} \pi_{\mu_3}^{\lambda_3}
\bigg(\varepsilon^{p_1\kappa\mu_1\mu_3}p_3^{\mu_2}\bigg)=\pi_{\mu_1}^{\lambda_1}
	\pi_{\mu_2}^{\lambda_2}  \pi_{\mu_3}^{\lambda_3}\bigg(-p_1^2\varepsilon^{\kappa\mu_1\mu_2\mu_3}+\varepsilon^{p_1\mu_1\mu_2\mu_3}p_1^\kappa-\varepsilon^{p_1\kappa\mu_1\mu_2}p_1^{\mu_3}\bigg)\\
	&\pi_{\mu_1}^{\lambda_1}
	\pi_{\mu_2}^{\lambda_2}  \pi_{\mu_3}^{\lambda_3}\bigg(\varepsilon^{p_1\kappa\mu_2\mu_3}p_2^{\mu_1}\bigg)=\pi_{\mu_1}^{\lambda_1}
	\pi_{\mu_2}^{\lambda_2} \pi_{\mu_3}^{\lambda_3}
	\bigg(\frac{1}{2}\left(p_1^2+p_2^2-p_3^2\right)\varepsilon^{\kappa\mu_1\mu_2\mu_3}+\varepsilon^{p_1\kappa\mu_1\mu_2}p_1^{\mu_3}+\varepsilon^{p_1\mu_1\mu_2\mu_3}p_2^\kappa\bigg).
\end{align}
More technical details and a full list of such tensor relations is given in the Appendix. Combining all the expressions we obtain  the projected equation
\begin{align}
	&\pi_{\mu_1}^{\lambda_1}(p_1)
	\pi_{\mu_2}^{\lambda_2} (p_2) \pi_{\mu_3}^{\lambda_3}
	(p_3) \bigg(\mathcal{K}^\kappa\braket{j^{\mu_1}(p_1)j^{\mu_2}(p_2)j_5^{\mu_3}(p_3)}\bigg)=\notag\\
	&=\pi_{\mu_1}^{\lambda_1}(p_1)
	\pi_{\mu_2}^{\lambda_2} (p_2) \pi_{\mu_3}^{\lambda_3}
	(p_3) \bigg[\,p_1^\kappa\bigg(\,C_{11}\,\varepsilon^{ \mu_1\mu_2\mu_3p_1}+\,C_{12}\,
	\varepsilon^{\mu_1\mu_2\mu_3p_2}+\,C_{13}\,
	\varepsilon^{\mu_1\mu_2p_1p_2}p_1^{\mu_3}\bigg)\notag\\
	&\hspace{1cm}+p_2^\kappa\bigg(\,C_{21}\,\varepsilon^{ \mu_1\mu_2\mu_3p_1}+\,C_{22}\,
	\varepsilon^{\mu_1\mu_2\mu_3p_2}+\,C_{23}\,
	\varepsilon^{\mu_1\mu_2p_1p_2}p_1^{\mu_3}\bigg)+C_{31}\varepsilon^{\kappa\mu_1\mu_2\mu_3}+C_{32}\varepsilon^{\kappa\mu_1\mu_2 p_1}p_1^{\mu_3}\notag\\
	&\hspace{4cm}+C_{33}\varepsilon^{\kappa\mu_1\mu_2 p_2}p_1^{\mu_3}+C_{34}\varepsilon^{\kappa\mu_1p_1p_2}\delta^{\mu_2\mu_3}+C_{35}\varepsilon^{\kappa\mu_2p_1p_2}\delta^{\mu_1\mu_3}+C_{36}\varepsilon^{\kappa\mu_3p_1p_2}\delta^{\mu_1\mu_2}
	\bigg],
\end{align}
where the $C_{ij}$ are scalar differential equations involving the form factors. In particular, $C_{1j}$ and $C_{2j}$ are of the second order, while all the others are first order differential equations. The action of $\mathcal{K}^\kappa$ on the longitudinal part is then given as
\begin{align}
	&\pi_{\mu_1}^{\lambda_1}(p_1)
	\pi_{\mu_2}^{\lambda_2} (p_2) \pi_{\mu_3}^{\lambda_3}
	(p_3) \bigg(\mathcal{K}^\kappa\braket{J^{\mu_1}(p_1)J^{\mu_2}(p_2)j_{5,loc}^{\mu_3}(p_3)}\bigg)=
	\pi_{\mu_1}^{\lambda_1}\left(p_1\right)
	\pi_{\mu_2}^{\lambda_2} \left(p_2\right) \pi_{\mu_3}^{\lambda_3}
	\left(p_3\right) 
	\Bigl[  \mathcal{A}\,\delta^{\mu_3\kappa} \varepsilon^{\mu_1\mu_2p_1p_2}
	\Bigr]\notag\\
	&=
	\pi_{\mu_1}^{\lambda_1}\left(p_1\right)
	\pi_{\mu_2}^{\lambda_2} \left(p_2\right) \pi_{\mu_3}^{\lambda_3}
	\left(p_3\right) 
	\Big[ \mathcal{A}\Big( \varepsilon^{\kappa\mu_2 p_1p_2}\delta^{\mu_1\mu_3}-\varepsilon^{\kappa\mu_1p_1p_2}\delta^{\mu_2\mu_3}+\varepsilon^{\kappa\mu_1\mu_2p_1}p_1^{\mu_3}+\varepsilon^{\kappa\mu_1\mu_2p_2}p_1^{\mu_3}\Big)\Big],
\end{align}
where 
\begin{equation}
	\mathcal{A}\equiv 	-\frac{16\, a \, i \, (\Delta_3-1)}{p_3^2}
\end{equation}
is related to the chiral anomaly. Due to the independence of the tensor structures listed above, the special conformal equations are written as
\begin{align}
&C_{ij}=0,\quad i=1,2,\quad j=1,2,3 	\label{eq:primff} \\ 
&C_{31}=0, \label{eq:secff1}\\
&C_{3j}+\mathcal{A}=0,\quad j=2,3,5 \label{eq:secff2} \\
&C_{34}-\mathcal{A}=0, \label{eq:secff3}\\
&C_{36}=0 \label{eq:secff4}
\end{align}
the explicit form of the primary equations \eqref{eq:primff} is
\begin{equation} \label{eq:PrimNonOmog}
\begin{aligned} 
&K_{31}\,A_1=0,\hspace{6cm} K_{32}\,A_1=0,\\
&K_{31}\,A_2=0,\hspace{6cm} K_{32}\,A_2=\left(\frac{4}{p_1^2}-\frac{2}{p_1}\frac{\partial}{\partial p_1}\right)A_2(p_1\leftrightarrow p_2)+2A_1,\\
&K_{31}\,A_2(p_1\leftrightarrow p_2)=\left(\frac{4}{p_2^2}-\frac{2}{p_2}\frac{\partial}{\partial p_2}\right)A_2-2A_1,\qquad K_{32}\,A_2(p_1\leftrightarrow p_2)=0,
\end{aligned}
\end{equation}
where we have defined
\begin{align}
K_i=\frac{\partial^2}{\partial p_i^2}+\frac{(d+1-2\Delta_i)}{p_i}\frac{\partial}{\partial p_i},\qquad K_{ij}=K_i-K_j.
\end{align}
These equations can also be reduced to a set of homogenous equations by repeatedly applying  the operator $K_{ij}$ and we have
\begin{equation} \label{eq:PrimOmog}
	\begin{aligned} 
		&K_{31}\,A_1=0,\qquad K_{32}\,A_1=0,\\
		&K_{31}\,A_2=0,\qquad K_{32}K_{32}\,A_2=0.
	\end{aligned}
\end{equation}

\subsection{Solutions of the CWIs}
The most general solution of the conformal Ward identities of the $VVA$ can be written in terms of integrals involving a product of three Bessel functions, namely 3K integrals \cite{Bzowski:2013sza}. For a detailed review on the properties of such integrals, see also \cite{Bzowski:2015yxv}.
We recall the definition of the general 3K integral
\begin{equation}
	I_{\alpha\left\{\beta_1 \beta_2 \beta_3\right\}}\left(p_1, p_2, p_3\right)=\int d x x^\alpha \prod_{j=1}^3 p_j^{\beta_j} K_{\beta_j}\left(p_j x\right)
\end{equation}
where $K_\nu$ is a modified Bessel function of the second kind 
\begin{equation}
	K_\nu(x)=\frac{\pi}{2} \frac{I_{-\nu}(x)-I_\nu(x)}{\sin (\nu \pi)}, \qquad \nu \notin \mathbb{Z} \qquad\qquad I_\nu(x)=\left(\frac{x}{2}\right)^\nu \sum_{k=0}^{\infty} \frac{1}{\Gamma(k+1) \Gamma(\nu+1+k)}\left(\frac{x}{2}\right)^{2 k}
\end{equation}
with the property
\begin{equation}
	K_n(x)=\lim _{\epsilon \rightarrow 0} K_{n+\epsilon}(x), \quad n \in \mathbb{Z}
\end{equation}
We will also use the reduced version of the triple-K integral defined as
\begin{equation}
	J_{N\left\{k_j\right\}}=I_{\frac{d}{2}-1+N\left\{\Delta_j-\frac{d}{2}+k_j\right\}}
\end{equation}
where we introduced the condensed notation $\{k_j \} = \{k_1, k_2, k_3 \}$.
The 3K integral satisfies an equation analogous to the dilatation equation with scaling degree \cite{Bzowski:2013sza}
\begin{equation}
	\text{deg}\left(J_{N\left\{k_j\right\}}\right)=\Delta_t+k_t-2 d-N
\end{equation}
where 
\begin{equation}
	k_t=k_1+k_2+k_3,\qquad\qquad \Delta_t=\Delta_1+\Delta_2+\Delta_3
\end{equation}
From this analysis, it is simple to relate the form factors to the triple-K integrals. Indeed, the dilatation Ward identities tell us that the form factor $A_i$ needs to be written as a combination of integrals of the following type
\begin{equation}
	J_{N_i+k_t,\{k_1,k_2,k_3\}}
\end{equation}
where $N_i$ is the number of momenta that the form factor multiplies in the decomposition \eqref{decompFin}. The special conformal Ward identities fix the remaining indices $k_1$, $k_2$ and $k_3$.
Indeed, recalling the following property of the 3K integrals
\begin{equation}
	K_{n m} J_{N\left\{k_j\right\}}=-2 k_n J_{N+1\left\{k_j-\delta_{j n}\right\}}+2 k_m J_{N+1\left\{k_j-\delta_{j m}\right\}}
\end{equation}
we can write the most general solution of the homogeneous equations \eqref{eq:PrimOmog} as
\begin{align}
A_1&=\alpha_1\,J_{3\{0,0,0\}},\\
A_2&=\alpha_{2}\,J_{1\{0,0,0\}}+\alpha_{3}\,J_{2\{0,1,0\}}.
\end{align}
Applying this general solutions to the non-homogenous equations \eqref{eq:PrimNonOmog}, we find the constraint
\begin{align}
\alpha_2=-2\alpha_3
\end{align}
and the solution reduces to 
\begin{align}
	A_1&=\alpha_1\,J_{3\{0,0,0\}},\\
	A_2&=-2\alpha_{3}\,J_{1\{0,0,0\}}+\alpha_{3}\,J_{2\{0,1,0\}}.
\end{align}
We now consider the first order differential equations \eqref{eq:secff1}, \eqref{eq:secff2}, \eqref{eq:secff3} and \eqref{eq:secff4} in their explicit form, ignoring the trivial ones.
We start with
\begin{align}
C_{36}&=\left(\frac{4}{p_2^2}-\frac{2}{p_2}\frac{\partial}{\partial p_2}\right)A_2-\left(\frac{4}{p_1^2}-\frac{2}{p_1}\frac{\partial}{\partial p_1}\right)A_2(p_1\leftrightarrow p_2)-2A_1=0
\end{align}
and using the property of the 3K integral
\begin{equation}
\frac{\partial}{\partial p_n}J_{N\{k_j\}}=-p_nJ_{N+1\{k_j-\delta_{jn}\}}
\end{equation}
we have
\begin{align}
C_{36}&=2\alpha_1\,J_{3\{0,0,0\}}=0,\notag\\
\end{align}
obtaining the constraint 
\begin{equation}
\alpha_1=0. 
\end{equation}
Taking into account this constraint, we consider the other secondary Ward identity as
\begin{align}
C_{31}&=2\,p_2\frac{\partial}{\partial p_2}\,A_2(p_1\leftrightarrow p_2)-2\,p_1\frac{\partial}{\partial p_1}\,A_2+\frac{2(p_1^2+p_2^2-p_3^2)}{p_1^2p_2^2}\bigg(p_1^2A_2-p_2^2\,A_2(p_1\leftrightarrow p_2)\bigg)+\notag\\
&\frac{2\,p_1\cdot p_2}{p_2}\frac{\partial}{\partial p_2}A_2-\frac{2\,p_1\cdot p_2}{p_1}\frac{\partial}{\partial p_1}A_2(p_1\leftrightarrow p_2)=0,
\end{align}
and this equation with the constraints obtained before on the coefficients $\alpha_i$ is trivially satisfied. 
The last equation we have to consider is the one related to the anomaly that takes the form
\begin{align}
	-\frac{2}{p_3^2}\left(p_2\frac{\partial}{\partial p_2}+p_1\frac{\partial}{\partial p_1}+2\right)\bigg(A_2+A_2(p_1\leftrightarrow p_2)\bigg)-\left(\frac{4}{p_1^2}-\frac{2}{p_1}\frac{\partial}{\partial p_1}\right)A_2(p_1\leftrightarrow p_2)-\mathcal{A}=0.
\end{align}
This equation can be easily solved by taking the limit $p_{3\mu}\rightarrow0$, $p_{1 \mu}=-p_{2 \mu}=p_\mu$. Assuming that $\alpha>\beta_t-1$ and $\beta_3>0$, we can write \cite{Bzowski:2013sza}
\begin{equation}
	\lim _{p_3 \rightarrow 0} I_{\alpha\left\{\beta_j\right\}}\left(p, -p, p_3\right)=p^{\beta_t-\alpha-1} \ell_{\alpha\left\{\beta_j\right\}}
\end{equation}
and then
\begin{equation}
	\lim _{p_3 \rightarrow 0} J_{N\{k_j\}}\left(p, -p, p_3\right)=\lim _{p_3 \rightarrow 0} I_{\frac{d}{2}-1+N\left\{\Delta_j-\frac{d}{2}+k_j\right\}}\left(p, -p, p_3\right)=p^{k_t+\Delta_3-2-N}\, \ell_{\frac{d}{2}-1+N\left\{\Delta_j-\frac{d}{2}+k_j\right\}},
\end{equation}
where
\begin{equation}
	\ell_{\alpha\left\{\beta_j\right\}}=\frac{2^{\alpha-3} \Gamma\left(\beta_3\right)}{\Gamma\left(\alpha-\beta_3+1\right)} \Gamma\left(\frac{\alpha+\beta_t+1}{2}-\beta_3\right) \Gamma\left(\frac{\alpha-\beta_t+1}{2}+\beta_1\right) \Gamma\left(\frac{\alpha-\beta_t+1}{2}+\beta_2\right) \Gamma\left(\frac{\alpha-\beta_t+1}{2}\right). 
\end{equation}
With this limit the equation can be solved and we have, with $d=4$ and $\Delta_3=3$, the constraint
\begin{align}
\alpha_3=8i\, a.
\end{align}
In summary, once the conformal constraints are solved, we find the solution of the transverse part in terms of one coefficient, proportional to the anomaly and in particular we have
	\begin{align}
		\braket{j^{\mu_1}(p_1)j^{\mu_2}(p_2) j^{\mu_3}_5 (p_3)} &=\,\pi^{\mu_1}_{\alpha_1}(p_1)
		\pi^{\mu_2}_{\alpha_2} (p_2) \pi^{\mu_3}_{\alpha_3}
		\left(p_3\right)\bigg[ 8ia
		\bigg(-2\,J_{1\{0,0,0\}}+\,J_{2\{0,1,0\}}\bigg)\, \varepsilon^{p_1 \alpha_1\alpha_2\alpha_3}\notag\\
		&\hspace{5cm}  -
		8ia
		\bigg(-2\,J_{1\{0,0,0\}}+\,J_{2\{1,0,0\}}\bigg)\, \varepsilon^{p_2\alpha_1\alpha_2\alpha_3}  
		\bigg]
	\end{align}
or in terms of the simplified version of the $3K$ integrals as
\begin{align}
	J_{1\{0,0,0\}}=I_{2\left\{1,1,1\right\}},\quad J_{2\{0,1,0\}}=I_{3\left\{1,2,1\right\}},\quad J_{2\{1,0,0\}}=I_{3\left\{2,1,1\right\}}.
\end{align}
Explicitly we have
	\begin{align}
	\braket{j^{\mu_1}(p_1)j^{\mu_2}(p_2) j^{\mu_3}_5 (p_3)} &=8ia\,\pi^{\mu_1}_{\alpha_1}(p_1)
	\pi^{\mu_2}_{\alpha_2} (p_2) \pi^{\mu_3}_{\alpha_3}
	\left(p_3\right)\bigg[ 
	\bigg(-2\,I_{2\left\{1,1,1\right\}}+\,I_{3\left\{1,2,1\right\}}\bigg)\, \varepsilon^{p_1 \alpha_1\alpha_2\alpha_3}\notag\\
	&\hspace{5cm}  -
	\bigg(-2\,I_{2\left\{1,1,1\right\}}+\,I_{3\left\{2,1,1\right\}}\bigg)\, \varepsilon^{p_2\alpha_1\alpha_2\alpha_3}  
	\bigg].
\label{bb1}
\end{align}
Furthermore, this expression can be reduced by noticing that
\begin{align}
-2\,I_{2\left\{1,1,1\right\}}+\,I_{3\left\{1,2,1\right\}}&=p_2^2\,I_{3\{1,0,1\}},\\
-2\,I_{2\left\{1,1,1\right\}}+\,I_{3\left\{1,2,1\right\}}&=p_1^2\,I_{3\{0,1,1\}},
\end{align}
finally giving for \eqref{bb1}
	\begin{align}
	\braket{j^{\mu_1}(p_1)j^{\mu_2}(p_2) j^{\mu_3}_5 (p_3)} &=8ia\,\pi^{\mu_1}_{\alpha_1}(p_1)
	\pi^{\mu_2}_{\alpha_2} (p_2) \pi^{\mu_3}_{\alpha_3}
	\left(p_3\right)\bigg[ \,p_2^2\,I_{3\{1,0,1\}}\, \varepsilon^{p_1 \alpha_1\alpha_2\alpha_3}-
	\,p_1^2\,I_{3\{0,1,1\}}\, \varepsilon^{p_2\alpha_1\alpha_2\alpha_3}  
	\bigg]
\end{align}

\section{Reducing the 3K integral in the solution}
Using the reduction relations presented in \cite{Bzowski:2015yxv, Bzowski:2020lip} we have that the solution is finite and can be reduced to the standard perturbation results as
\begin{align}
&I_{3\left\{1,0,1\right\}}=p_1\,p_3\frac{\partial^2}{\partial p_1\partial p_3}\,I_{1\{0,0,0\}}
\end{align}
where 
$I_{1\{0,0,0\}}$ is the master integral related to the three-point function of the operator $\varphi^2$ in the theory of free massless scalar $\varphi$ in $d=4$. Indeed, the relation between the master integrals and the massless scalar $1$-loop $3$-point momentum space integral is
\begin{align}
	I_{1\{0,0,0\}}=(2\pi)^2K_{4,\{1,1,1\}}=(2\pi)^2\int\frac{d^4k}{(2\pi)^4}\frac{1}{k^2\,(k-p_1)^2\,(k+p_2)^2}=\frac{1}{4}C_{0}(p_1^2,p_2^2,p_3^2)
\end{align}
where 
\begin{align}
	K_{d\{\delta_1\delta_2\delta_3\}}=\int\frac{d^dk}{(2\pi)^d}\frac{1}{(k^2)^{\delta_3}\,((k-p_1)^2)^{\delta_2}\,((k+p_2)^2)^{\delta_1}}.
\end{align}
By using the relations of the derivative acting on the master integral presented in \cite{Coriano:2018bbe}, and analytically continuing 
$C_0$ to $d$ dimensions we find that
\begin{align}
	p_3\frac{\partial}{\partial p_3}C_{0}(p_1^2,p_2^2,p_3^2)=&\frac{1}{\lambda}\bigg\{2(d-3) \bigg[(p_1^2-p_2^2+p_3^2)B_0(p_1^2)+(p_2^2+p_3^2-p_1^2)B_0(p_2^2)-2p_3^2B_0(p_3^2)\bigg]\notag\\
	&+\bigg[(d-4)(p_1^2-p_2^2)^2-(d-2)p_3^4+2p_3^2(p_1^2+p_2^2)\bigg]C_0(p_1^2,p_2^2,p_3^2)\bigg\}
\end{align}
where $\lambda$ is the K\"allen $\lambda$-function 
\begin{align}
\lambda\equiv\left(p_1-p_2-p_3\right) \left(p_1+p_2-p_3\right) \left(p_1-p_2+p_3\right) \left(p_1+p_2+p_3\right).
\end{align}
Then
\begin{align}
&4 \, I_{3\left\{1,0,1\right\}}=p_1 \frac{\partial}{\partial p_1}\frac{1}{\lambda}\bigg\{2(d-3) \bigg[(p_1^2-p_2^2+p_3^2)B_0(p_1^2)+(p_2^2+p_3^2-p_1^2)B_0(p_2^2)-2p_3^2B_0(p_3^2)\bigg]\notag\\
&\hspace{2cm}+\bigg[(d-4)(p_1^2-p_2^2)^2-(d-2)p_3^4+2p_3^2(p_1^2+p_2^2)\bigg]C_0(p_1^2,p_2^2,p_3^2)\bigg\}\notag\\
&=-\frac{4p_1^2(p_1^2-p_2^2-p_3^2)}{\lambda^2}\bigg\{2(d-3) \bigg[(p_1^2-p_2^2+p_3^2)B_0(p_1^2)+(p_2^2+p_3^2-p_1^2)B_0(p_2^2)-2p_3^2B_0(p_3^2)\bigg]\notag\\
&\hspace{2cm}+\bigg[-(d-2)p_3^4+2p_3^2(p_1^2+p_2^2)\bigg]C_0(p_1^2,p_2^2,p_3^2)\bigg\}\notag\\
&+\frac{1}{\lambda}\bigg\{2(d-3)\bigg[2p_1^2\,B_0(p_1^2)+(p_1^2-p_2^2+p_3^2)\,(d-4)B_0(p_1^2)-2p_1^2\,B_0(p_2^2)\bigg]+4p_1^2p_3^2\,C_0(p_1^2,p_2^2,p_3^2)\notag\\
&+\frac{1}{\lambda}\bigg[-(d-2)p_3^2+2p_3^2(p_1^2+p_2^2)\bigg]\bigg[2(d-3)\bigg((p_1^2+p_2^2-p_3^2)B_0(p_2^2)+(p_1^2-p_2^2+p_3^2)B_0(p_3^2)-2p_1^2B_0(p_1^2)\bigg)\notag\\
&\qquad+\Big(-(d-2)p_1^4+2p_1^2(p_2^2+p_3^2)\Big)C_0(p_1^2,p_2^2,p_3^2)
\bigg]
\bigg\}+O(d-4).
\end{align}
This expression is finite in the expansion around $d=4$ and the result simplifies to the form
\begin{align}
	I_{3\left\{1,0,1\right\}}(p_1^2,p_2^2,p_3^2)=&\frac{1}{\lambda^2}\bigg\{-2 p_1^2 p_3^2 \bigg[p_1^2 \left(p_2^2-2 p_3^2\right)+p_1^4+p_2^2 p_3^2-2 p_2^4+p_3^4\bigg] C_0\left(p_1^2,p_2^2,p_3^2\right)\notag\\
	&+p_1^2 \left(\left(p_1^2-p_2^2\right)^2+4p_2^2p_3^2- p_3^4\right)\log\left(\frac{p_1^2}{p_2^2}\right)+4p_1^2 p_3^2 \left(p_1^2-p_3^2\right)\log\left(\frac{p_1^2}{p_3^2}\right)\notag\\
	&-p_3^2 \Big((p_2^2-p_3^2)^2+4p_1^2p_2^2-p_1^4\Big)\log\left(\frac{p_2^2}{p_3^2}\right)-\lambda(p_1^2-p_2^2+p_3^2)\bigg\}
\end{align}
with the form factor given by
\begin{align}
	A_2^{(CFT)}(p_1,p_2,p_3)=8ia\,p_2^2\,I_{3\{1,0,1\}}(p_1^2,p_2^2,p_3^2).
\end{align}
\subsection{Perturbative realization of the correlator}
The same approach that we have investigated in the previous section, can be redone in the context of a free field theory, with a single chiral fermion. The steps are the same as above, with the correlator expanded in the same basis of form factors identified above using \eqref{decomp} and \eqref{decomp1}, using a free Dirac fermion 
and the only form factor $A_2$ is given by
\begin{align}
	A_2^{(P)}=&\frac{-e^3\,p_2^2}{2\pi^2\lambda^2}\bigg\{-\lambda\,(p_1^2-p_2^2+p_3^2)+p_1^2 \bigg[\left(p_1^2-p_2^2\right)^2+4 p_3^2 p_2^2-p_3^4\bigg] \log \left(\frac{p_1^2}{p_2^2}\right)\notag\\
	&+p_3^2 \bigg[p_1^4-4 p_1^2 p_2^2-\left(p_2^2-p_3^2\right)^2\bigg] \log \left(\frac{p_2^2}{p_3^2}\right)+4p_1^2p_3^2\Big(p_1^2-p_3^2\Big)\log \left(\frac{p_1^2}{p_3^2}\right)\notag\\
	&-2 p_1^2 p_3^2 \bigg[p_1^2 \left(p_2^2-2 p_3^2\right)+p_1^4+p_2^2 p_3^2-2 p_2^4+p_3^4\bigg] C_0\left(p_1^2,p_2^2,p_3^2\right)\bigg\}.
\end{align}
On can directly check the complete match between the perturbative and the CFT result, once  the anomaly coefficient is chosen of the form 
\begin{align}
a=\frac{i\,e^3}{16 \pi^2}.
\end{align}

\section{The conformal $\langle AAA \rangle $ correlator }
In this section we illustrate how conformal invariance completely determines the structure of the $\langle J_5J_5J_5\rangle$ correlator. Differently from the $\langle JJJ_5\rangle$ correlator, a 3K integral regularization is needed in this case. Therefore, we will work in $d=4+\epsilon$ and then perform the limit $\epsilon\rightarrow0$.

\subsection{Longitudinal/Transverse decomposition}
First of all, we consider the anomalous Ward identity 
\begin{equation}
	\nabla_\mu J_5^\mu=a' \varepsilon^{\mu \nu \rho \sigma} F^A_{\mu \nu} F^A_{\rho \sigma}
\end{equation}
where $F_{\mu\nu}^A$ is the gauge field coupled to the axial current.
We impose such identity symmetrically on the all the three external axial-vector currents of the $AAA$ correlator, leading to the following equations
\begin{equation}
	\begin{aligned}
		&p_{1 \mu_1}	\left\langle J_5^{\mu_1 }(p_1) J_5^{\mu_2 } (p_2)J_5^{\mu_3}(p_3)\right\rangle=-8 \, a' \, i \, \varepsilon^{p_1p_2\mu_2\mu_3}\\
		&p_{2 \mu_2}	\left\langle J_5^{\mu_1 }(p_1) J_5^{\mu_2 } (p_2)J_5^{\mu_3}(p_3)\right\rangle=8 \, a' \, i \, \varepsilon^{p_1p_2\mu_1\mu_3}\\
		&p_{3 \mu_3}	\left\langle J_5^{\mu_1 }(p_1) J_5^{\mu_2 } (p_2)J_5^{\mu_3}(p_3)\right\rangle=-8 \, a' \, i \, \varepsilon^{p_1p_2\mu_1\mu_2}.
	\end{aligned}
\end{equation}
Notice that the equations above are symmetric in the three momenta, but for technical reasons we prefer to express them only in terms of $p_1$ and $p_2$.
Note that, if we contract the correlator with multiple momenta at the same time, the result is zero.\\
We can then decompose the correlator into the following transverse and longitudinal parts
\begin{equation}
	\left\langle J_5^{\mu_1 } J_5^{\mu_2 } J_5^{\mu_3}\right\rangle=\left\langle j_5^{\mu_1 } j_5^{\mu_2 } j_5^{\mu_3}\right\rangle
	+\left\langle j_{5 \text { loc }}^{\mu_1 } j_5^{\mu_2 } j_5^{\mu_3}\right\rangle
	+\left\langle j_5^{\mu_1 } j_{5 \text { loc }}^{\mu_2 } j_5^{\mu_3}\right\rangle
	+\left\langle j_5^{\mu_1 } j_5^{\mu_2 } j_{5 \text { loc }}^{\mu_3}\right\rangle
\end{equation}
The longitudinal parts are completely fixed by the anomalous Ward identity above in the form
\begin{equation}
	\begin{aligned}
		&\left\langle J^{\mu_1}_{5\text { loc}}\left(p_1\right) J_5^{\mu_2}\left(p_2\right) j_{5}^{\mu_3}\left(p_3\right)\right\rangle=-\frac{8 \, a' i}{p_1^2} \varepsilon^{p_1 p_2 \mu_2 \mu_3} p_1^{\mu_1} \\
		&\left\langle J_5^{\mu_1}\left(p_1\right) J_{5\text { loc}}^{\mu_2}\left(p_2\right) j_{5 }^{\mu_3}\left(p_3\right)\right\rangle=\frac{8\, a' i}{p_2^2} \varepsilon^{p_1 p_2 \mu_1 \mu_3} p_2^{\mu_2} \\
		&\left\langle J_5^{\mu_1}\left(p_1\right) J_5^{\mu_2}\left(p_2\right) j_{5 \text { loc}}^{\mu_3}\left(p_3\right)\right\rangle=-\frac{8\, a' i}{p_3^2} \varepsilon^{p_1 p_2 \mu_1 \mu_2} p_3^{\mu_3}. 
	\end{aligned}
\end{equation}
On the other hand, the transverse part can be expressed as
\begin{equation}
	\begin{aligned}
		\left\langle j_5^{\mu_1}\left(p_1\right) 
		j_5^{\mu_2}\left(p_2\right) j^{\mu_3}_5 	\left(p_3\right)\right\rangle &=\pi^{\mu_1}_{\alpha_1}\left(p_1\right)
		\pi^{\mu_2}_{\alpha_2} \left(p_2\right) \pi^{\mu_3}_{\alpha_3}
		\left(p_3\right)\Bigl[  \\
		\tilde{A}(p_1,p_2,p_3)& \varepsilon^{p_1p_2\alpha_1\alpha_2}p_1^{\alpha_3} + 
		A(p_1,p_2,p_3) \varepsilon^{p_1 \alpha_1\alpha_2\alpha_3}  -
		A(p_2,p_1,p_3) \varepsilon^{p_2\alpha_1\alpha_2\alpha_3} 
		\Bigr].
	\end{aligned}
\end{equation}
Differently from the $\langle JJJ_5 \rangle$ correlator, there is an additional Bose symmetry condition we have to consider: $\{p_1,\mu_1\} \leftrightarrow \{p_3,\mu_3\}$. Therefore, in this case, there is only one independent form factor. Indeed we have
\begin{equation}
	\tilde{A}(p_1,p_2,p_3)=2\, \frac{A(p_1,p_2,p_3)+A(p_2,p_3,p_1)+A(p_3,p_1,p_2)}{p_1^2+p_2^2+p_3^2}
\end{equation}
and moreover
\begin{equation}
	{A}(p_1,p_2,p_3)=\frac{(p_1^2-p_2^2+p_3^2) A(p_3, p_2, p_1) - 2 p_2^2 A(p_1, p_3, p_2) - 2 p_2^2 A(p_2, p_1, p_3)  }{p_1^2 + p_2^2 + p_3^2}
\end{equation}
For now on we will ignore such relations and we will treat $\tilde{A}$ and $A$ as independent quantities. We can still check later that our final result is invariant under the exchange of the currents.

\subsection{Dilatation and Special Conformal Ward identities}
The dilatation Ward identities of the transverse part are not affected by the longitudinal terms. Therefore the constraints are the same of the $\langle JJJ_5 \rangle$ correlator
\begin{equation}
	\begin{aligned}
		&\sum_{i=1}^3 p_i \frac{\partial \tilde{A}}{\partial p_i}-\left(\Delta_3-5\right) \tilde{A}=0\\
		&\sum_{i=1}^3 p_i \frac{\partial A}{\partial p_i}-\left(\Delta_3-3\right) A=0.
	\end{aligned}
\end{equation}
The invariance of the correlator with respect to the special conformal transformations is encoded in the following relation
\begin{equation}
	\begin{aligned}
		0=\pi_{\mu_1}^{\lambda_1}\left(p_1\right)
		\pi_{\mu_2}^{\lambda_2} \left(p_2\right) \pi_{\mu_3}^{\lambda_3}&
		\left(p_3\right) \mathcal{K}^k 
		\Biggl[ 
		\left\langle j_5^{\mu_1 } j_5^{\mu_2 } j_5^{\mu_3}\right\rangle
		+\left\langle j_{5 \text { loc }}^{\mu_1 } j_5^{\mu_2 } j_5^{\mu_3}\right\rangle
		+\left\langle j_5^{\mu_1 } j_{5 \text { loc }}^{\mu_2 } j_5^{\mu_3}\right\rangle
		+\left\langle j_5^{\mu_1 } j_5^{\mu_2 } j_{5 \text { loc }}^{\mu_3}\right\rangle
		\Biggr].
	\end{aligned}
\end{equation}
We procede in a manner similar to the $\langle JJJ_5\rangle$ correlator, using the same Schouten identities.
We can then decompose the contribution of the transverse part into the following form factors
\begin{equation}
	\begin{aligned}
		&\pi_{\mu_1}^{\lambda_1}(p_1)
		\pi_{\mu_2}^{\lambda_2} (p_2) \pi_{\mu_3}^{\lambda_3}
		(p_3) \bigg(\mathcal{K}^\kappa\langle j^{\mu_1}(p_1)j^{\mu_2}(p_2)j_5^{\mu_3}(p_3)\rangle\bigg)=\notag\\
		&=\pi_{\mu_1}^{\lambda_1}(p_1)
		\pi_{\mu_2}^{\lambda_2} (p_2) \pi_{\mu_3}^{\lambda_3}
		(p_3) \bigg[\,p_1^\kappa\bigg(\,C_{11}\,\varepsilon^{ \mu_1\mu_2\mu_3p_1}+\,C_{12}\,
		\varepsilon^{\mu_1\mu_2\mu_3p_2}+\,C_{13}\,
		\varepsilon^{\mu_1\mu_2p_1p_2}p_1^{\mu_3}\bigg)\notag\\
		&\hspace{1cm}+p_2^\kappa\bigg(\,C_{21}\,\varepsilon^{ \mu_1\mu_2\mu_3p_1}+\,C_{22}\,
		\varepsilon^{\mu_1\mu_2\mu_3p_2}+\,C_{23}\,	\varepsilon^{\mu_1\mu_2p_1p_2}p_1^{\mu_3}\bigg)+C_{31}\varepsilon^{\kappa\mu_1\mu_2\mu_3}+C_{32}\varepsilon^{\kappa\mu_1\mu_2 p_1}p_1^{\mu_3}\notag\\
		&\hspace{4cm}+C_{33}\varepsilon^{\kappa\mu_1\mu_2 p_2}p_1^{\mu_3}+C_{34}\varepsilon^{\kappa\mu_1p_1p_2}\delta^{\mu_2\mu_3}+C_{35}\varepsilon^{\kappa\mu_2p_1p_2}\delta^{\mu_1\mu_3}+C_{36}\varepsilon^{\kappa\mu_3p_1p_2}\delta^{\mu_1\mu_2}
		\bigg],
	\end{aligned}
\end{equation}
where the explicit expression for the form factor is the same of the $\langle JJJ_5\rangle$.
However, both the primary and secondary equations will contain anomalous terms from the longitudinal parts of the correlators. Indeed, the primary Ward identities are given by
\begin{equation} \label{eq:secSpecEqAAA}
	\begin{aligned}
		&C_{11}= 0 \hspace{4.5cm}
		&&C_{21}=-\frac{16(d-2) a' i}{p_1^2}\\
		&C_{12}=\frac{16(d-2) a' i}{p_2^2} 
		&&C_{22}=0\\
		&C_{13}=0 
		&&C_{23}=0
	\end{aligned}
\end{equation}
while the secondary equations can be written as
\begin{equation} \label{eq:secSpecEqAAA}
	\begin{aligned}
		&C_{31}=-\frac{8 i a' (d-2) (p_1-p_2) (p_1+p_2) \left(p_1^2+p_2^2-p_3^2\right)}{p_1^2 p_2^2}\qquad\qquad
		&&C_{32}=\frac{32 i a'}{p_3^2}-\frac{16 i a' (d-2)}{p_1^2}\\
		&C_{33}=\frac{32 i a'}{p_3^2}-\frac{16 i a' (d-2)}{p_2^2} \qquad
		&&C_{34}=- \frac{32  i a'}{p_3^2}+\frac{16 i a' (d-2)}{p_2^2}\\
		&C_{35}=-\frac{16 i a' (d-2)}{p_1^2}+\frac{32 i a'}{p_3^2} \qquad
		&&C_{36}=-\frac{16 i a' (d-2) (p_1-p_2) (p_1+p_2)}{p_1^2 p_2^2}
	\end{aligned}
\end{equation}
The explicit form of the primary special conformal Ward identity is 
\begin{equation}
	\begin{aligned}
		&K_{31}\,\tilde{A}=0, \hspace{4.5cm}  K_{32}\,\tilde{A}=0,\\
		&K_{31}\,A=0, \hspace{4.5cm}   K_{32}\,A= 2 \left( \frac{d-2}{p_1^2}-\frac{1}{p_1}\frac{\partial}{\partial p_1}\right)A(p_1\leftrightarrow p_2)+2\tilde{A}+\frac{16a' i\left(d-2 \right)}{p_1^2}\\
		& K_{31}\,A(p_1\leftrightarrow p_2)= 2 \left( \frac{d-2}{p_2^2}-\frac{1}{p_2}\frac{\partial}{\partial p_2}\right)A-2\tilde{A}+\frac{16a' i\left(d-2 \right)}{p_2^2},\hspace{2cm}
		K_{32}\,A(p_1\leftrightarrow p_2)=0
	\end{aligned}
\end{equation}
These equations can also be reduced to a set of homogenous equations by repeatedly applying the operator $K_{ij}$
\begin{align}
	&K_{31}\,\tilde{A}=0,&& K_{32}\,\tilde{A}=0,\\
	&K_{31}\,A=0,&& K_{32}K_{32}\,A=0,\\
	&K_{32}\,A(p_1\leftrightarrow p_2)=0,\quad && K_{31}K_{31}\,A(p_1\leftrightarrow p_2)=0
\end{align}
As we can see, the presence of anomalous terms in the primary equations does not affect the structure of the homogenous equations which are exactly the same of the $\langle JJJ_5\rangle$ correlator. 

\subsection{Solutions of the CWIs}
The solutions of the primary homogeneous equations can be written in terms of the following 3K integrals, extending the previous approach of the $VVA$
\begin{align}
	\tilde{A}&=\alpha_1\,J_{3\{0,0,0\}}=\alpha_1 I_{ \frac{d}{2}+2\{ \frac{d}{2}-1 , \frac{d}{2}-1, \frac{d}{2}-1 \}},\\
	A&=\alpha_{2}\,J_{1\{0,0,0\}}+\alpha_{3}\,J_{2\{0,1,0\}}=\alpha_{2}
	I_{\frac{d}{2}\{ \frac{d}{2} -1 ,\frac{d}{2} -1 ,\frac{d}{2} -1 \}}
	+\alpha_{3} I_{\frac{d}{2}+1 ,\{\frac{d}{2} -1,\frac{d}{2} , \frac{d}{2}-1 \}}. 
\end{align}
Inserting our solutions back to the non-homogeneous equations, in the limit $p_3\rightarrow0$ we find
\begin{equation}
	\left(\frac{d-4}{3}\right)\alpha_1= { -\frac{i a' 2^{6-\frac{d}{2}} (d-2) \sin \left(\frac{\pi  d}{2}\right)}{\pi  \Gamma \left(\frac{d}{2}+1\right)}+{\alpha_2}+{\alpha_3} (d-2)}.
\end{equation}
We then focus on the first two secondary equations in \eqref{eq:secSpecEqAAA}.
The explicit form of the first equation is
\begin{equation}
	\begin{aligned}
		0=&\frac{p_1^2+p_2^2-p_3^2 }{p_1^2 \, p_2^2}\Bigg\{-p_1^2 p_2^2 {\tilde{A}}+(d-2) \Big[ p_1^2 A-p_2^2 A(p_2\leftrightarrow p_1)+8 i a' (p_1^2-p_2^2) \Big] \Bigg\}\\
		&+\left(p_1^2+p_2^2-p_3^2\right) \left(
		\frac{1 }{p_1}\frac{\partial}{\partial p_1}A(p_2\leftrightarrow p_1)-\frac{1 }{p_2}\frac{\partial}{\partial p_2}A\right)
		-2 p_1 \frac{\partial}{\partial p_1}A
		+2 p_2 \frac{\partial}{\partial p_2}A(p_2\leftrightarrow p_1)
	\end{aligned}
\end{equation}
which leads to the condition
\begin{equation}
	{\alpha_2}= (d-2) \left(-{\alpha_3}+\frac{i a' 2^{6-\frac{d}{2}}  \sin \left(\frac{\pi  d}{2}\right)}{\pi  \Gamma \left(\frac{d}{2}+1\right)}\right).
\end{equation}
The explicit form of the second equation in \eqref{eq:secSpecEqAAA} is
\begin{equation}
	\begin{aligned}
		0=-2\, &\frac{2 p_1^2+ (d-2) p_3^2}{p_1^2 p_3^2}A(p_2\leftrightarrow p_1)
		+2 \left(1-\frac{p_1^2}{p_3^2}\right)\frac{1}{p_1}  \frac{\partial}{\partial p_1}A(p_2\leftrightarrow p_1) 
		-\frac{2 p_1 }{p_3^2}\frac{\partial}{\partial p_1}A\\&
		-\frac{2 p_2 }{p_3^2}\frac{\partial}{\partial p_2}A
		-\frac{2 p_2 }{p_3^2}\frac{\partial}{\partial p_2}A(p_2 \leftrightarrow p_1)		
		-\frac{4 A}{p_3^2}+\frac{p_1 \left(-p_1^2+p_2^2+p_3^2\right) }{p_3^2}
		\frac{\partial}{\partial p_1}\tilde{A} \\&
		+\frac{p_2 \left(-p_1^2+p_2^2-3 p_3^2\right) }{p_3^2}\frac{\partial}{\partial p_2}\tilde{A}
		-\frac{2 \left(2 p_1^2-2 p_2^2+p_3^2\right) \tilde{A}}{p_3^2}-16 \,a'\, i\, \left(\frac{d-2}{ p_1^2}-\frac{2}{p_3^2}\right)
	\end{aligned}
\end{equation}
which leads to the constraint
\begin{equation}
	{\alpha_3}= \frac{i \, a' \, 2^{7-\frac{d}{2}} (d-1) \sin \left(\frac{\pi  d}{2}\right)}{\pi  (d-4) \Gamma \left(\frac{d}{2}+1\right)}.
\end{equation}
The other secondary equations don't impose any other constraints.
We insert such conditions into our solution and use the following property of the 3K integral
\begin{equation}
	{I}_{\frac{d}{2}+1\{\frac{d}{2}-1,\frac{d}{2},\frac{d}{2}-1 \}}= p_2^2 {I}_{\frac{d}{2}+1\{\frac{d}{2}-1,\frac{d}{2}-2,\frac{d}{2}-1 \}}+(d-2)  {I}_{\frac{d}{2}\{\frac{d}{2}-1,\frac{d}{2}-1,\frac{d}{2}-1 \}}          
\end{equation}
in order to arrive to the following expression in $d=4+\epsilon$
\begin{equation}
	\begin{aligned}
		&\tilde{A}=0,\\
		&A=24\,i\,a' p_2^2  I_{3\{1,0,1\}}+8\,i \, a' \epsilon \,  I_{2+\frac{\epsilon}{2}  \{1+\frac{\epsilon}{2} ,1+\frac{\epsilon}{2} ,1+\frac{\epsilon}{2} \}}.
	\end{aligned}
\end{equation}
Note that we are keeping the second term of $A$ because the 3K integral has a pole in $\epsilon$.
In the end we have
\begin{equation}\label{eq:AAAresult}
	\begin{aligned}
		\langle j_5^{\mu_1}\left(p_1\right) &
		j_5^{\mu_2}\left(p_2\right) j_5^{\mu_3} 	\left(p_3\right)\rangle =  \pi^{\mu_1}_{\alpha_1}\left(p_1\right)
		\pi^{\mu_2}_{\alpha_2} \left(p_2\right) \pi^{\mu_3}_{\alpha_3}
		\left(p_3\right) \, 24 \, i \, a' \,  \Bigl[ \\&
		\left(
		I_{3\{1,0,1\}}\, p_2^2
		+\frac{\epsilon}{3} \,  I_{2+\frac{\epsilon}{2}  \{1+\frac{\epsilon}{2} ,1+\frac{\epsilon}{2} ,1+\frac{\epsilon}{2} \}}
		\right) \varepsilon^{p_1 \alpha_1\alpha_2\alpha_3}  -
		\left(
		I_{3\{0,1,1\}}\, p_1^2
		+\frac{\epsilon}{3} \,  I_{2+\frac{\epsilon}{2}  \{1+\frac{\epsilon}{2} ,1+\frac{\epsilon}{2} ,1+\frac{\epsilon}{2} \}}
		\right) \varepsilon^{p_2 \alpha_1\alpha_2\alpha_3} 
		\Bigr]
	\end{aligned}.
\end{equation}
\subsection{Connection with the $\langle VVA\rangle$ correlator}
Recalling that $a'=a/3$, our results are in accordance with the formula
\begin{equation}
	\left\langle j_5^{\mu_1 } j_5^{\mu_2 } j_5^{\mu_3}\right\rangle=\frac{1}{3}\left(
	\left\langle j_5^{\mu_1 } j^{\mu_2 } j^{\mu_3}\right\rangle+
	\left\langle j^{\mu_1 } j_5^{\mu_2 } j^{\mu_3}\right\rangle+
	\left\langle j^{\mu_1 } j^{\mu_2 } j_5^{\mu_3}\right\rangle
	\right)
\end{equation}
In order to prove such formula we use the relation
\begin{equation}
	\begin{aligned}
		\left\langle j^{\mu_1}\left(p_1\right) 
		j^{\mu_2}\left(p_2\right) j^{\mu_3}_5 	\left(p_3\right)\right\rangle =  \pi^{\mu_1}_{\alpha_1}\left(p_1\right)
		\pi^{\mu_2}_{\alpha_2} \left(p_2\right) \pi^{\mu_3}_{\alpha_3}
		\left(p_3\right) \, 24 \, i \, a' \,  \Bigl[  
		I_{3\{1,0,1\}}\, p_2^2 \, \varepsilon^{p_1 \alpha_1\alpha_2\alpha_3}  -
		I_{3\{0,1,1\}}\, p_1^2 \, \varepsilon^{p_2\alpha_1\alpha_2\alpha_3}  
		\Bigr]
	\end{aligned}.
\end{equation}
At this stage, exchanging the second current with the third, we have
\begin{equation}
	\begin{aligned}
		\left\langle j^{\mu_1}\left(p_1\right) 
		j_5^{\mu_2}\left(p_2\right) j^{\mu_3}	\left(p_3\right)\right\rangle = & \pi^{\mu_1}_{\alpha_1}\left(p_1\right)
		\pi^{\mu_2}_{\alpha_2} \left(p_2\right) \pi^{\mu_3}_{\alpha_3}
		\left(p_3\right) \, 24 \, i \, a' \,  \Bigl[  \\&
		- \left(p_1^2\, I_{3\{0,1,1\}}+p_3^2\, I_{3\{1,1,0\}}\right)\varepsilon^{p_1 \alpha_1\alpha_2\alpha_3}
		- p_1^2\, I_{3\{0,1,1\}}\,\varepsilon^{p_2 \alpha_1\alpha_2\alpha_3},
		\Bigr]
	\end{aligned}
\end{equation}
while, exchanging the first current with the third one, we obtain
\begin{equation}
	\begin{aligned}
		\left\langle j_5^{\mu_1}\left(p_1\right) 
		j^{\mu_2}\left(p_2\right) j^{\mu_3}	\left(p_3\right)\right\rangle =  & \pi^{\mu_1}_{\alpha_1}\left(p_1\right)
		\pi^{\mu_2}_{\alpha_2} \left(p_2\right) \pi^{\mu_3}_{\alpha_3}
		\left(p_3\right) \, 24 \, i \, a' \,  \Bigl[  \\&
		p_2^2 \, I_{3\{1,0,1\}} \, \varepsilon^{p_1 \alpha_1\alpha_2\alpha_3}+
		\left(p_2^2\, I_{3\{1,0,1\}}+p_3^2\, I_{3\{1,1,0\}}\right)\varepsilon^{p_2 \alpha_1\alpha_2\alpha_3}
		\Bigr]
	\end{aligned}
\end{equation}
Summing the last three equations and using the following identity of the 3K integrals
\begin{equation}
	p_1^2I_{3\{0,1,1\}}+
	p_2^2I_{3\{1,0,1\}}+
	p_3^2I_{3\{1,1,0\}}
	=-\epsilon \, I_{2+\frac{\epsilon}{2}\{1+\frac{\epsilon}{2},1+\frac{\epsilon}{2},1+\frac{\epsilon}{2}\}}
\end{equation}
we arrive at Eq. \eqref{eq:AAAresult}. 

\section{Comparison with other parameterizations}
Having established the agreement between the perturbative (lowest order) and the non-perturbative computation of the correlator, we try to relate the result of the expansion introduced in \eqref{decomp1} with the two most popular parameterizations of the same vertex. \\
As we have already mentioned in the introduction, the original parameterization of the $VVA$ was presented in \cite{Rosenberg:1962pp}. Lorentz symmetry and parity fix the correlation function in the form 
\begin{align}
\langle J^{\mu_1}(p_1)J^{\mu_2}(p_2)J^{\mu_3}_5(p_3) \rangle &= B_1 (p_1, p_2) \varepsilon^{p_1\mu_1\mu_2\mu_3} + B _2 (p_1, p_2)\varepsilon^{p_2\mu_1\mu_2\mu_3} +
B_3 (p_1, p_2) \varepsilon^{p_1p_2\mu_1\mu_3}{p_1}^{\mu_2} \nonumber \\
&+  B_4 (p_1, p_2) \varepsilon^{p_1p_2\mu_1\mu_3}p_2^{\mu_2}
+ B_5 (p_1, p_2)\varepsilon^{p_1p_2\mu_2\mu_3}p_1^{\mu_1}
+ B_6 (p_1, p_2) \varepsilon^{p_1p_2\mu_2\mu_3}p_2^{\mu_1},\nonumber \\
\label{Ross}
\end{align}
with $B_1$ and $B_2$ divergent by power counting.
If we use a diagrammatic evaluation of the correlator, the four invariant amplitudes $B_i$ for $i\geq3$ are given by explicit parametric integrals \cite{Rosenberg:1962pp}
\begin{align}
\label{ssym}
B_3(p_1, p_2) &= - B_6 (p_2, p_1) =   16 \pi^2 I_{11}(p_1, p_2), \nonumber \\
B_4(p_1,p_2) &= - B_5 (p_2, p_1) =- 16 \pi^2 \left[ I_{20}(p_1,p_2) - I_{10}(p_1,p_2) \right],
\end{align}
where the general massive $I_{st}$ integral is defined by
\begin{equation}
I_{st}(p_1,p_2) = \int_0^1 dw \int_0^{1-w} dz w^s z^t \left[ z(1-z) p_1^2 + w(1-w) p_2^2 + 2 w z (p_1\cdot p_2) - m^2 \right]^{-1}.
\end{equation}
Both $B_1$ and $B_2$ can be rendered finite by imposing the Ward identities on the two vector lines, giving
\begin{align}
B_1 (p_1,p_2) &= p_1 \cdot p_2 \, B_3 (p_1,p_2) + p_2^2 \, B_4 (p_1,p_2),
\label{WI1} \\
B_2 (p_1,p_2) &= p_1^2 \, B_5 (p_1,p_2) + p_1 \cdot p_2 \, B_6 (p_1,p_2),
\label{WI2}
\end{align}
which allow to re-express the formally divergent amplitudes in terms of the convergent ones. 
The Bose symmetry on the two vector vertices is fulfilled thanks to the relations
\begin{equation}
	\begin{aligned}
		B_5\left(p_1, p_2\right) & =-B_4\left(p_2, p_1\right) \\
		B_6\left(p_1, p_2\right) & =-B_3\left(p_2, p_1\right) .
	\end{aligned}
\end{equation}
Using the conservation WIs for the vector currents, we obtain the convergent expansion \cite{Armillis:2009sm}
\begin{align}
\langle J^{\mu_1}J^{\mu_2}&J^{\mu_3}_5 \rangle= B_3 (p_1 \cdot p_2 \varepsilon^{p_1\mu_1\mu_2\mu_3} + p_1^{\mu_2} \varepsilon^{p_1p_2\mu_1 \mu_3}) +
B_4 (p_2 \cdot p_2 \varepsilon^{p_1\mu_1\mu_2\mu_3} + p_2^{\mu_2} \varepsilon^{p_1p_2\mu_1 \mu_3} ) \notag \\
&+ B_5 (p_1 \cdot p_1 \varepsilon^{p_2\mu_1\mu_2\mu_3}+ p_1^{\mu_1} \varepsilon^{p_1p_2\mu_2\mu_3}) +
B_6 (p_1 \cdot p_2 \varepsilon^{p_2\mu_1\mu_2\mu_3} + p_2^{\mu_1} \varepsilon^{p_1p_2\mu_2\mu_3} ) \notag\\
& \equiv B_3 \, \eta_3^{\mu_1\mu_2\mu_3}(p_1,p_2) + B_4 \, \eta_4^{\mu_1\mu_2\mu_3}(p_1,p_2)
+ B_5 \, \eta_5^{\mu_1\mu_2\mu_3}(p_1,p_2) + B_6 \, \eta_6^{\mu_1\mu_2\mu_3}(p_1,p_2),
\label{reduced}
\end{align}
\begin{table}[t]
\begin{center}
\begin{tabular}
{|c |c |}   \hline
$\eta_1$ & $ p_1^{\mu_3} \, \varepsilon^{p_1p_2\mu_1\mu_2} $  \\ \hline
$\eta_2$ & $ p_2^{\mu_3} \, \varepsilon^{p_1p_2\mu_1\mu_2}$ \\ \hline\hline
$\eta_3$         &           $p_1 \cdot p_2 \varepsilon^{p_1\mu_1\mu_2\mu_3} + p_1^{\mu_2} \varepsilon^{p_1p_2\mu_1 \mu_3}$ \\ \hline
$\eta_4$ & $p_2 \cdot p_2 \varepsilon^{p_1\mu_1\mu_2\mu_3} + p_2^{\mu_2} \varepsilon^{p_1p_2\mu_1 \mu_3} $ \\ \hline
$\eta_5$ & $ p_1 \cdot p_1 \varepsilon^{p_2\mu_1\mu_2\mu_3}+ p_1^{\mu_1} \varepsilon^{p_1p_2\mu_2\mu_3} $  \\ \hline
$\eta_6$ & $p_1 \cdot p_2 \varepsilon^{p_2\mu_1\mu_2\mu_3} + p_2^{\mu_1} \varepsilon^{p_1p_2\mu_2\mu_3} $  \\ \hline
\end{tabular}
\caption{ Tensor structures of odd parity in the expansion of the $VVA$ with conserved vector currents. \label{table2} }
\end{center}
\end{table}
where in the last step we have introduced four tensor structures that are mapped into one another under the Bose symmetry of the two vector lines. One can identifies six of them, as indicated in Table 1, but two of them\begin{align}
\eta_1^{\mu_1\mu_2\mu_3}(p_1,p_2) =  p_1 ^{\mu_3} \, \varepsilon^{p_1p_2\mu_1\mu_2},  \qquad \qquad
\eta_2^{\mu_1\mu_2\mu_3}(p_1,p_2) =  p_2 ^{\mu_3} \, \varepsilon^{p_1p_2\mu_1 \mu_2},
\end{align}
 are related by the Schouten relations to the other four, $\eta_3,\ldots \eta_6$. Indeed one has 
\begin{align}
\eta_1 ^{\mu_1\mu_2\mu_3}(p_1,p_2) &= \eta_3 ^{\mu_1\mu_2\mu_3}(p_1,p_2) - \eta_5 ^{\mu_1\mu_2\mu_3}(p_1,p_2),    \\
\eta_2 ^{\mu_1\mu_2\mu_3}(p_1,p_2) &= \eta_4 ^{\mu_1\mu_2\mu_3}(p_1,p_2) - \eta_6 ^{\mu_1\mu_2\mu_3}(p_1,p_2).
\end{align}
The remaining tensor structures are inter-related by the Bose symmetry
\beq
\eta^{\mu_1\mu_2\mu_3}_3(p_1,p_2)=-\eta^{\mu_2\mu_1\mu_3}_6(p_2,p_1)  \qquad \eta_4^{\mu_1\mu_2\mu_3}(p_1,p_2)=-\eta_5^{\mu_2\mu_1\mu_3}(p_2,p_1). 
\eeq
The correct counting of the independent form factors/tensor structures can be done only after we split each of them into their symmetric and antisymmetric components 
\beq
\eta^{\mu_1\mu_2\mu_3}_i=\eta_i^{S \, \, \mu_1\mu_2\mu_3} +\eta_i^{A \, \, \mu_1\mu_2\mu_3}\qquad  \eta_i^{S/A \, \, \mu_1\mu_2\mu_3}\equiv\frac{1}{2}\left( \eta^{\mu_1\mu_2\mu_3}_i(p_1,p_2) \pm \eta^{ \mu_2\mu_1\mu_3}_i(p_2,p_1)\right)\equiv \eta_i^{\pm \, \, \mu_1\mu_2\mu_3}
\eeq
with $i\geq 3$, giving 
\beq
\eta_3^+(p_1,p_2)=-\eta_6^+(p_1,p_2) \qquad \eta_3^-(p_1,p_2)=\eta_6^-(p_1,p_2) 
\eeq
\beq
\eta_4^+(p_1,p_2)=-\eta_5^+(p_1,p_2) \qquad \eta_4^-(p_1,p_2)=\eta_5^-(p_1,p_2). 
\eeq
where we omitted all the tensorial indices which are in the order $\mu_1 \, \mu_2 \, \mu_3$.
We can then re-express the correlator as
\beq
\langle VVA \rangle=B_3^+\eta_3^+ + B_3^-\eta_3^- +B_4^+\eta_4^+ + B_4^-\eta_4^-
\eeq
 in terms of four tensor structures of definite symmetry times 4 independent form factors. 

\subsection{L/T decomposition} 
An alternative parameterization of the $VVA$ correlator, which allows to set a direct comparison with the one that we have introduced in the previous sections is given by \cite{Knecht:2003xy}
\beq
\langle J^{\mu_1}(p_1)J^{\mu_2}(p_2)J^{\mu_3}_5(p_3) \rangle=\frac{1}{8\pi^2} \left({W^{L}}^{\mu_1\mu_2\mu_3} - {  \mathcal \, W^{T}}^{\mu_1\mu_2\mu_3} \right)
\eeq
where the longitudinal component is specified in eq. \eqref{refe}, while
the transverse component is given by
\begin{align}
	{  \mathcal \, W^{T}}^{\mu_1\mu_2\mu_3}(p_1,p_2,p_3^2) &=
	w_T^{(+)}\left(p_1^2, p_2^2,p_3^2 \right)\,t^{(+)\,\mu_1\mu_2\mu_3}(p_1,p_2)
	+\,w_T^{(-)}\left(p_1^2,p_2^2,p_3^2\right)\,t^{(-)\,\mu_1\mu_2\mu_3}(p_1,p_2) \nonumber \\
	& +\,\, {\widetilde{w}}_T^{(-)}\left(p_1^2, p_2^2,p_3^2 \right)\,{\widetilde{t}}^{(-)\,\mu_1\mu_2\mu_3}(p_1,p_2),
\end{align}
This decomposition automatically account for all the symmetries of the correlator
with the transverse tensors given by
\begin{equation}\label{tensors}
	\begin{aligned}
		t^{(+) \, \mu_1\mu_2\mu_3}(p_1,p_2) &=
		p_{1}^{\mu_2}\, \veps^{ \mu_1\mu_3 p_1p_2}  -
		p_{2}^{\mu_1}\,\veps^{\mu_2\mu_3 p_1 p_2}  - (p_{1} \cdot p_2)\,\veps^{\mu_1\mu_2\mu_3(p_1 - p_2)}\\ &\hspace{4cm}
		+  \frac{p_1^2 + p_2^2 - p_3^2}{p_3^2}\,  (p_1+p_2)^{\mu_3} \, 
		\veps^{\mu_1\mu_2 p_1 p_2}
		\nonumber  , \\
		t^{(-)\,\mu_1\mu_2\mu_3}(p_1,p_2) &= \left[ (p_1 - p_2)^{\mu_3} - \frac{p_1^2 - p_2^2}{p_3^2}\, (p_{1}+p_2)^{ \mu_3} \right] \,\veps^{\mu_1\mu_2 p_1 p_2}
		\nonumber\\
		{\widetilde{t}}^{(-)\, \mu_1\mu_2\mu_3}(p_1,p_2) &= p_{1}^{\mu_2}\,\veps^{ \mu_1\mu_3 p_1p_2} +
		p_{2}^{\mu_1}\,\veps^{\mu_2\mu_3 p_1 p_2} 
		- (p_{1}\cdot p_2)\,\veps^{ \mu_1 \mu_2 \mu_3 (p_1+p_2)}.
	\end{aligned}
\end{equation}
The map between the Rosenberg representation and the current one is given by the relations 
\begin{align}
B_3 (p_1, p_2) &= \frac{1}{8 \pi^2} \left[ w_L - \tilde{w}_T^{(-)}
-\frac{p_1^2+p_2^2}{p_3^2}     w_T^{(+)}
- 2 \,  \frac{p_1 \cdot p_2 + p_2^2 }{p_3^2 }w_T^{(-)}  \right],  \\
B_4 (p_1, p_2) &= \frac{1}{8 \pi^2} \left[  w_L
+ 2 \, \frac{p_1 \cdot p_2}{p_3^2}       w_T^{(+)}
+ 2 \, \frac{p_1 \cdot p_2 + p_1^2}{k^2 }w_T^{(-)}  \right], 
\end{align}
and viceversa
\begin{equation}
\label{ppo}
w_L (p_1^2, \, p_2^2,p_3^2) = \frac{8 \pi^2}{p_3^2} \left[B_1 - B_2 \right]
\end{equation}
or, after the imposition of the Ward identities in Eqs.(\ref{WI1},\ref{WI2}),
\begin{align}
w_L ( p_1^2, \, p_2^2,p_3^2) &= \frac{8 \pi^2}{p_3^2}
\left[ (B_3-B_6) p_1 \cdot p_2 + B_4 \, p_2^2 - B_5 \, p_1^2 \right],
\label{wL}\\
w_T^{(+)} (p_1^2, \, p_2^2,p_3^2)  &= - 4 \pi^2 \left(B_3 - B_4 + B_5 - B_6 \right),
\label{wTp}\\
w_T^{(-)} ( p_1^2, \, p_2^2,p_3^2)  &=  4 \pi^2 \left(B_4+B_5 \right),
\label{wTm}\\
\tilde{w}_T^{(-)} ( p_1^2, \, p_2^2,p_3^2)  &= - 4\pi^2 \left( B_3 + B_4 + B_5 + B_6 \right),
\label{wTt}
\end{align}
where $B_i\equiv B_i(p_1,p_2)$. As already mentioned, \eqref{ppo} is a special relation, since it shows that the pole is not affected by Chern-Simons forms, telling us of the physical character of this part of the interaction. \\
Also in this case, the counting of the form factor is four, one for the longitudinal pole part and 3 for the transverse part. Notice that all of them are either symmetric or antisymmetric by construction. 
\beqa
w_L(p_1^2,p_2^2,p_3^2)&=& w_L(p_2^2,p_1^2,p_3^2) \nn\\
w_T^{(+)}(p_1^2,p_2^2,p_3^2)&=& w_T^{(+)}(p_2^2,p_1^2,p_3^2) \nn\\
w_T^{(-)}(p_1^2,p_2^2,p_3^2)&=& - w_T^{(-)}(p_2^2,p_1^2,p_3^2) \nn \\
\tilde{w}_T^{(-)}(p_1^2,p_2^2,p_3^2)&=&- \tilde{w}_T^{(-)}(p_2^2,p_1^2,p_3^2). 
\eeqa
To relate this decomposition to our,  we apply the transverse projectors and obtain 
\begin{align}
	&\pi_{\mu_1}^{\lambda_1}(p_1)
	\pi_{\mu_2}^{\lambda_2} (p_2) \pi_{\mu_3}^{\lambda_3}
	(p_3) \bigg(W^{T\,\mu_1\mu_2\mu_3}(p_1,p_2,p_3)\bigg)=\notag\\
	&=\pi_{\mu_1}^{\lambda_1}(p_1)
	\pi_{\mu_2}^{\lambda_2} (p_2) \pi_{\mu_3}^{\lambda_3}
	(p_3)\bigg[\big(w_T^{(+)}+{\widetilde{w}}_T^{(-)}\big)p_3^{\alpha_2}\, \varepsilon^{\alpha_1\alpha_3p_1p_2}-2\,w_T^{(-)}\,p_1^{\alpha_3} \,\varepsilon^{\alpha_1\alpha_2p_1p_2}\notag\\
	&\qquad+\big(w_T^{(+)}-{\widetilde{w}}_T^{(-)}\big)p_2^{\alpha_1}\, \varepsilon^{\alpha_2\alpha_3p_1p_2}+\big(w_T^{(+)}+{\widetilde{w}}_T^{(-)}\big)(p_1\cdot p_2)\varepsilon^{\alpha_1\alpha_2\alpha_3p_1}-\big(w_T^{(+)}-{\widetilde{w}}_T^{(-)}\big)(p_1\cdot p_2)\varepsilon^{\alpha_1\alpha_2\alpha_3p_2}\bigg],
\end{align}
and by using the Schouten identities
\begin{align}
	\pi^{\mu_1}_{\alpha_1}
	\pi^{\mu_2}_{\alpha_2} \pi^{\mu_3}_{\alpha_3}\bigg(p_2^{\alpha_1} \varepsilon^{p_1 p_2 {\alpha_2} {\alpha_3}}\bigg)&=\pi^{\mu_1}_{\alpha_1}
	\pi^{\mu_2}_{\alpha_2} \pi^{\mu_3}_{\alpha_3}\bigg(-\big(p_2\cdot p_1\big) \varepsilon^{p_2\alpha_1 \alpha_2\alpha_3}+p_2^2\,\varepsilon^{p_1\alpha_1\alpha_2\alpha_3}+p_1^{\alpha_3}\,\varepsilon^{ p_1p_2\alpha_1\alpha_2}\bigg),\\
	\pi^{\mu_1}_{\alpha_1}
	\pi^{\mu_2}_{\alpha_2} \pi^{\mu_3}_{\alpha_3}\bigg(p_3^{\alpha_2}\varepsilon^{p_1 p_2 \alpha_1\alpha_3}\bigg)&=\pi^{\mu_1}_{\alpha_1}
	\pi^{\mu_2}_{\alpha_2} \pi^{\mu_3}_{\alpha_3}\bigg(-p_1^2\,\varepsilon^{p_2 \alpha_1\alpha_2\alpha_3}+\big(p_1\cdot p_2\big)\varepsilon^{p_1\alpha_1\alpha_2\alpha_3}-p_1^{\alpha_3}\varepsilon^{p_1p_2\alpha_1\alpha_2}\bigg),
\end{align}
we obtain
\begin{align}
	\pi_{\mu_1}^{\lambda_1}(p_1)
	\pi_{\mu_2}^{\lambda_2} (p_2) \pi_{\mu_3}^{\lambda_3}
	(p_3) \bigg(W^{T\,\mu_1\mu_2\mu_3}(p_1,p_2,p_3)\bigg)&=\pi_{\mu_1}^{\lambda_1}(p_1)
	\pi_{\mu_2}^{\lambda_2} (p_2) \pi_{\mu_3}^{\lambda_3}
	(p_3)\bigg\{-p_1^2\big(w_T^{(+)}+{\widetilde{w}}_T^{(-)}\big)\varepsilon^{p_2 \alpha_1\alpha_2\alpha_3}\notag\\
	&+p_2^2\big(w_T^{(+)}-{\widetilde{w}}_T^{(-)}\big)\varepsilon^{p_1\alpha_1\alpha_2\alpha_3}-2\big({\widetilde{w}}_T^{(-)}+\,w_T^{(-)}\big)p_1^{\alpha_3}\,\varepsilon^{ p_1p_2\alpha_1\alpha_2}\bigg\}.
\end{align}
We then identify the form factors of our decomposition and the current L/T one in the form 
\begin{align}
	A_1&=\frac{1}{4 \pi^2} \big({\widetilde{w}}_T^{(-)}+\,w_T^{(-)}\big),\notag\\
	A_2&=-\frac{1}{8 \pi^2}\, p_2^2\big(w_T^{(+)}-{\widetilde{w}}_T^{(-)}\big).
\end{align}
Notice that $A_1$ is antisymmetric in the exchange of the two vector lines and counts for one independent form factor, while $A_2$ contains both symmetric and antisymmetric components and counts as two. Combined with $w_L$, we again find that our form factors are four in the general case, before enforcing the conformal WIs on the parameterization. One can check from the solution of the special CWIs that this number is reduced by one in both representations, since in this case   
\begin{equation}
	{\widetilde{w}}_T^{(-)}=-\,w_T^{(-)}
\end{equation}
for the L/T one. In our case the form factor $A_1$ vanishes
\begin{equation}
	A_1=0,
\end{equation}
and we are left with three form factors in both cases. 
Proceeding in a similar manner, we can also map the Rosenberg parametrization into the one we worked in. The results are
\begin{equation}
	\begin{aligned}
		&A_1=B_3-B_6\\
		&A_2=p_2^2(B_6  + B_4).
	\end{aligned}
\end{equation}
\section{Nonrenormalization theorems}
In this section we establish a connection between the conformal solutions of correlators of mixed chirality and the perturbative results previously derived for them in the chiral limit of QCD.
They take the form of non-renormalization theorems, originally presented in \cite{Knecht:2003xy}. \\
They are a direct consequence of the fact that correlators of currents of different chiralities vanish in such limit. Also in this case one identifies constraints between $w_L$, the pole part, and combinations of  transverse form factors, which are not affected by radiative corrections. \\
From the perturbative picture, these theorems can be understood quite easily at diagrammatic level. The reason lays in the absence of explicit mass insertions in the perturbative expansion of a $\langle J_LJ J_R \rangle $ correlator, if computed in the chiral limit. In this case, chirality flips on the fermion lines of the the contour of the diagram in Fig. \eqref{ddiag} are prohibited, and the vector-like nature of the QCD interactions with the fermions does the rest, guaranteeing the vanishing of the correlator. \\
As just mentioned, the theorems in  \cite{Knecht:2003xy} are derived from the 
$\langle J_LJ J_R \rangle $ vertex, where $J_L$ and $J_R$ are left and right chiral currents 
 \beq
 J_L\equiv \frac{1}{2}\left(J - J_5\right), \qquad    J_R\equiv \frac{1}{2}\left(J + J_5 \right)
 \eeq
 while $J$ is vector-like. The building block of $\langle J_LJ J_R \rangle $ is the $JJJ_5$ (or $VVA$) correlator.
All the other diagrams, in the chiral limit, such as the $AVV$ or $VAV$ or $AAA$, are trivially related to the $VVA$ due to the anticommuting property of $\gamma_5$ and the symmetry constrain
\beq
AAA=\frac{1}{3}\left( AVV + VAV + VVA\right)
\eeq
as illustrated in Fig. 2.
In perturbation theory, the anomaly ($a'$) content of the $AAA$, for instance, can be determined on the basis of symmetry, assuming an equal sharing ($a/3$) of the anomaly for each external axial-vector line. The constraint can be used as a starting point for moving the anomaly around the vertices, by the inclusion of appropriate Chern-Simons terms. In the Rosenberg representation \cite{Rosenberg:1962pp} they amount to shifts of the form factors $B_1$ and $B_2$ (see also the appendix).\\
In \cite{Knecht:2003xy}, the authors analyzed the $\langle J_L J J_R\rangle$ correlator 
\begin{equation}
	\langle J_L J J_R \rangle=\frac{1}{4}\Big[\langle VVV\rangle-\langle AVV\rangle-\langle AVA\rangle+\langle VVA\rangle\Big].
\end{equation}
Using the charge conjugation invariance, they set the parity-even contribution $\langle VVV\rangle$ and $\langle AVA\rangle$ to zero. Alternatively, we can assume the correlator is conformally invariant and arrive to the same conclusion \cite{Bzowski:2013sza}. We start from this point.\\
Conformal invariance requires the abelian even-parity correlator $\langle JJJ\rangle$ to be zero in any dimension. In order to show that, first we impose the following conservation Ward identities
\begin{equation}
	\begin{aligned}
		p_{i\mu_i}\,\braket{J^{\mu_1}(p_1)J^{\mu_2}(p_2)J^{\mu_3}(p_3)}=0,\quad \quad i=1,2,3 
	\end{aligned}
\end{equation}
Therefore the correlator is purely transverse and can be expressed as
\begin{equation}
	\begin{aligned}
		\left\langle J^{\mu_1 }\left({p}_1\right) J^{\mu_2 }\left({p}_2\right) J^{\mu_3 }\left({p}_3\right)\right\rangle& =\pi_{\alpha_1}^{\mu_1}\left({p}_1\right) \pi_{\alpha_2}^{\mu_2}\left({p}_2\right) \pi_{\alpha_3}^{\mu_3}\left({p}_3\right)\left[A_1(p_1,p_2,p_3) \, p_2^{\alpha_1} p_3^{\alpha_2} p_1^{\alpha_3}\right. \\
		+A_2 (p_1,p_2,p_3)& \, \delta^{\alpha_1 \alpha_2} p_1^{\alpha_3}+A_2\left(p_3, p_1, p_2\right) \delta^{\alpha_1 \alpha_3} p_3^{\alpha_2} 
		\left.+A_2 \left(p_2, p_3, p_1\right) \delta^{\alpha_2 \alpha_3} p_2^{\alpha_1}\right] .
	\end{aligned}
\end{equation}
The $A_1$ form factor is completely antisymmetric for any permutation of the momenta while the form factor $A_2$ is antisymmetric under $(p_1\leftrightarrow p_2)$. We now consider the conformal constraints on the form factors.
The dilatation Ward identities are
\begin{equation}
	\begin{aligned}
		&\sum_{i=1}^{3} p_i \frac{\partial A_1}{\partial p_i }-\left(d-6\right) A_1=0,\\
		&\sum_{i=1}^{3} p_i \frac{\partial A_2}{\partial p_i }-\left(d-4\right) A_2=0,\\
	\end{aligned}
\end{equation}
The primary special conformal Ward identities are
\begin{equation}
	\begin{aligned}
		&{K}_{12} A_1=0, \qquad\qquad && {K}_{13} A_1=0, \\
		&{K}_{12} A_2=0, && {K}_{13} A_2=2 A_1.
	\end{aligned}
\end{equation}
The solution to such equations can be written in terms of the following 3K integrals
\begin{equation} \label{eq:solVVVomog}
	\begin{aligned}
		& A_1=\alpha_1 J_{3\{000\}} \\
		& A_2=\alpha_1 J_{2\{001\}}+\alpha_2 J_{1\{000\}}
	\end{aligned}
\end{equation}
The secondary special conformal Ward identities are 
\begin{equation}
	\begin{aligned}
		&L_3[A_1(p_1,p_2,p_3)]+2R[A_2(p_1,p_2,p_3)-A_2(p_3,p_1,p_2)]=0\\
		&L_1[A_2(p_2,p_3,p_1)]+2p_1^2A_2(p_3,p_1,p_2)-2p_1^2A_2(p_1,p_2,p_3)=0
	\end{aligned}
\end{equation}
where we defined the following operators
\begin{equation}
	\begin{aligned}
		\mathrm{L}_N & =p_1\left(p_1^2+p_2^2-p_3^2\right) \frac{\partial}{\partial p_1}+2 p_1^2 p_2 \frac{\partial}{\partial p_2} \\
		& +\left[\left(2 d-\Delta_1-2 \Delta_2+N\right) p_1^2+\left(2 \Delta_1-d\right)\left(p_3^2-p_2^2\right)\right] \\
		\mathrm{R} & =p_1 \frac{\partial}{\partial p_1}-\left(2 \Delta_1-d\right)
	\end{aligned}
\end{equation}
Inserting our solution \eqref{eq:solVVVomog} into such equations, we arrive to the conditions $\alpha_1=\alpha_2=0$.\\
Since $\langle AVA\rangle$ is not anomalous, one can prove in the same way that this correlator vanishes too.
Therefore, we are left with 
\begin{equation}
	\langle J_L J J_R \rangle=\frac{1}{4}\Big( \langle VVA\rangle -\langle AVV\rangle\Big).
\end{equation}
The authors of \cite{Knecht:2003xy} assumed that the $\langle J_L J J_R \rangle$ correlator is simply a Chern-Simons term, in order to prove the nonrenormalization theorems. Using our conformal solution, we can directly write the expression for the $\langle J_L J J_R \rangle$ correlator and prove their statement. Indeed for the longitudinal part we can write
\begin{equation}
	\langle VVA\rangle_{loc}-\langle AVV\rangle_{loc} =-8 \,i \,a \bigg[ \frac{p_3^{\mu_3}}{p_3^2}\varepsilon^{p_1 p_2\mu_1\mu_2} -
	\frac{p_1^{\mu_1}}{p_1^2}\varepsilon^{p_1 p_2\mu_2\mu_3} 
	\bigg]
\end{equation}
while for the transverse part, after contracting the projectors in our solution, we have
\begin{equation}
	\begin{gathered}
		\langle VVA\rangle_{transv}-\langle AVV\rangle_{transv} =8 \,i \,a \, \epsilon \, I_{2+\frac{\epsilon}{2},\{1+\frac{\epsilon}{2},1+\frac{\epsilon}{2},1+\frac{\epsilon}{2}\}}
		\bigg[
		\varepsilon^{p_2\mu_1\mu_2\mu_3}-
		\frac{p_3^{\mu_3}}{p_3^2}\varepsilon^{p_1 p_2\mu_1\mu_2} +
		\frac{p_1^{\mu_1}}{p_1^2}\varepsilon^{p_1 p_2\mu_2\mu_3} 
		\bigg]\\=-
		8 \,i \,a 
		\bigg[
		\varepsilon^{p_2\mu_1\mu_2\mu_3}-
		\frac{p_3^{\mu_3}}{p_3^2}\varepsilon^{p_1 p_2\mu_1\mu_2} +
		\frac{p_1^{\mu_1}}{p_1^2}\varepsilon^{p_1 p_2\mu_2\mu_3} 
		\bigg]
	\end{gathered}
\end{equation}
where in the last row we used the explicit expression of the 3K integral.
Adding the contributions together, we arrive to 
\begin{equation}
	\langle J_L J J_R \rangle=\langle VVA\rangle-\langle AVV\rangle =-8 \,i \,a \, \varepsilon^{p_2\mu_1\mu_2\mu_3}.
\end{equation}
which tells us that the $\langle J_L J J_R \rangle$ correlator is simply given by a Chern-Simons term, as expected. Note that, proceeding in a similar manner, one can prove eq. \eqref{eq:VVAAAACS} too.\\
Of course, one can also check that our conformal solution directly satisfies the following nonrenormalization theorems, originally derived in the chiral limit of perturbative QCD
\begin{equation}
	0=\Bigg\{ 
	\Big[
	w_T^{(+)}+w_T^{(-)}	\Big]\left(q_1^2,q_2^2,\left(q_1+q_2\right)^2\right)-
	\Big[
	w_T^{(+)}+w_T^{(-)}	\Big]\left(\left(q_1+q_2\right)^2,q_2^2,q_1^2\right)
	 \Bigg\}_{\text{pQCD}}
\end{equation}
\begin{equation}
	0=\Bigg\{ 
	\Big[
	\tilde{w}_T^{(-)}+w_T^{(-)}	\Big]\left(q_1^2,q_2^2,\left(q_1+q_2\right)^2\right)+
	\Big[
	\tilde{w}_T^{(-)}+w_T^{(-)}	\Big]\left(\left(q_1+q_2\right)^2,q_2^2,q_1^2\right)
	\Bigg\}_{\text{pQCD}}
\end{equation}
and
\begin{equation}
	\begin{aligned}
		\Bigg\{ 
		\Big[
		w_T^{(+)}+\tilde{w}_T^{(-)}	\Big] & \left(q_1^2,q_2^2,\left(q_1+q_2\right)^2\right)+
		\Big[
		w_T^{(+)}+\tilde{w}_T^{(-)}	\Big]\left(\left(q_1+q_2\right)^2,q_2^2,q_1^2\right)
		\Bigg\}_{\text{pQCD}}-w_L\left(\left(q_1+q_2\right)^2,q_2^2,q_1^2\right)\\
		=&-\Bigg\{ 
		\frac{2(q_2^2+q_1\cdot q_2)}{q_1^2} w_T^{(+)}\left(\left(q_1+q_2\right)^2,q_2^2,q_1^2\right)
		-2\frac{q_1 \cdot q_2}{q_1^2}w_T^{(-)}\left(\left(q_1+q_2\right)^2,q_2^2,q_1^2\right)
		\Bigg\}_{\text{pQCD}}
	\end{aligned}
\end{equation}

\section{Conclusions} 
In this work we have illustrated how the CWIs in momentum space can be used to determine the structure of chiral anomaly diagrams in an autonomous way respect to coordinate space. This shows that anomalies in CFT can  be treated consistently in this specific framework, that allows to establish a link between such correlators and  the ordinary perturbative amplitudes. Parity-odd correlators are important in many physical context, and in this case we have shown how the conformal properties of previous perturbative analysis, performed in free field theory in the chiral limit, are the result of conformal symmetry and of its constraints. Our derivation does not rely on any Lagrangian realization. \\
The analysis of parity odd correlators, especially for 4-point functions, 
is still under investigation, given its complexity, and the inclusion of the anomaly content is for certainly an interesting aspect that deserves a closer attention. 
It may be possible in the future to consider mixed conformal/chiral anomalies in the same framework. We hope to come back to the investigation of these points in future work. 

\centerline{\bf Acknowledgements}
This work is partially supported by INFN within the Iniziativa Specifica QFT-HEP.  
The work of C. C. and S.L. is funded by the European Union, Next Generation EU, PNRR project "National Centre for HPC, Big Data and Quantum Computing", project code CN00000013 and by INFN iniziativa specifica QFT-HEP. 
M. M. M. is supported by the European Research Council (ERC) under the European Union as Horizon 2020 research and innovation program (grant agreement No818066) and by Deutsche Forschungsgemeinschaft (DFG, German Research Foundation) under Germany's Excellence Strategy EXC-2181/1 - 390900948 (the Heidelberg STRUCTURES Cluster of Excellence). We thank Raffaele Marotta, Pietro Santorelli, Massimo Taronna, Mario Cret\`i and Riccardo Tommasi for discussions. C.C. thanks the theory group of the Physics Department of the University "Federico II" of Naples for hospitality during the completion of this study.

\appendix
\section{Schouten identities\label{appendixA}}
In this section we derive the minimal decomposition to describe $X^{\kappa\mu_1\mu_2\mu_3}$ in eq. \eqref{kjjj}. We start by writing all the possible tensor structures   
\begin{align}
		&
		\varepsilon^{ \mu_1\mu_2\mu_3p_1}p_1^\kappa,\quad
		\varepsilon^{\mu_1\mu_2\mu_3p_2}p_1^\kappa,\quad
		\varepsilon^{\mu_1\mu_2p_1p_2}p_1^{\mu_3}p_1^\kappa,\quad
		\varepsilon^{\mu_1\mu_3p_1p_2}p_3^{\mu_2}p_1^\kappa,\quad
		\varepsilon^{\mu_2\mu_3p_1p_2}p_2^{\mu_1}p_1^\kappa,\nonumber\\[0.5ex]
		&\varepsilon^{ \mu_1\mu_2\mu_3p_1}p_2^\kappa,\quad
		\varepsilon^{\mu_1\mu_2\mu_3p_2}p_2^\kappa,\quad
		\varepsilon^{\mu_1\mu_2p_1p_2}p_1^{\mu_3}p_2^\kappa,\quad
		\varepsilon^{\mu_1\mu_3p_1p_2}p_3^{\mu_2}p_2^\kappa,\quad
		\varepsilon^{\mu_2\mu_3p_1p_2}p_2^{\mu_1}p_2^\kappa,\nonumber\\[2ex]
		&
		\varepsilon^{\mu_2\mu_3p_1p_2}\delta^{\mu_1\kappa},\quad
		\varepsilon^{\mu_1\mu_3p_1p_2}\delta^{\mu_2\kappa},\quad
		\varepsilon^{\mu_1\mu_2p_1p_2}\delta^{\mu_3\kappa},\nonumber\\[2ex]
		&
		\varepsilon^{\kappa\mu_1\mu_2\mu_3},\quad
		\varepsilon^{\kappa\mu_1\mu_2 p_1}p_1^{\mu_3},\quad
		\varepsilon^{\kappa\mu_1\mu_2 p_2}p_1^{\mu_3},\quad
		\varepsilon^{\kappa\mu_1\mu_3p_1}p_3^{\mu_2} ,\quad
		\varepsilon^{\kappa\mu_1\mu_3p_2}p_3^{\mu_2} ,\quad
		\varepsilon^{\kappa\mu_2\mu_3p_1}p_2^{\mu_1} ,\quad
		\varepsilon^{\kappa\mu_2\mu_3p_2}p_2^{\mu_1} , \nonumber\\[0.5ex]
		&\varepsilon^{\kappa\mu_1p_1p_2}\delta^{\mu_2\mu_3},\quad
		\varepsilon^{\kappa\mu_1p_1p_2}p_3^{\mu_2}p_1^{\mu_3} ,\quad
		\varepsilon^{\kappa\mu_2p_1p_2}\delta^{\mu_1\mu_3},\quad
		\varepsilon^{\kappa\mu_2p_1p_2}p_2^{\mu_1}p_1^{\mu_3} ,\quad
		\varepsilon^{\kappa\mu_3p_1p_2}\delta^{\mu_1\mu_2},\quad
		\varepsilon^{\kappa\mu_3p_1p_2}p_2^{\mu_1}p_3^{\mu_2} .
	\end{align}
Such tensor structures are not all independent, indeed we are going to show what are the Schouten identites one has to consider in order to find the minimal number of tensor structures. The first two identites are 
\begin{align}
\varepsilon^{[\mu_2\mu_3\kappa p_1}\delta^{\mu_1]\alpha}&=0,\\
\varepsilon^{[\mu_2\mu_3\kappa p_2}\delta^{\mu_1]\alpha}&=0,
\end{align}
that can be contracted with $p_{1\alpha}$ and $p_{2\alpha}$ and taking the projectors in front we obtain the four tensor identities
\begin{align}
	&\pi_{\mu_1}^{\lambda_1}
	\pi_{\mu_2}^{\lambda_2} \pi_{\mu_3}^{\lambda_3}
\bigg(\varepsilon^{p_1\kappa\mu_1\mu_3}p_3^{\mu_2}\bigg)=\pi_{\mu_1}^{\lambda_1}
	\pi_{\mu_2}^{\lambda_2}  \pi_{\mu_3}^{\lambda_3}\bigg(-p_1^2\varepsilon^{\kappa\mu_1\mu_2\mu_3}+\varepsilon^{p_1\mu_1\mu_2\mu_3}p_1^\kappa-\varepsilon^{p_1\kappa\mu_1\mu_2}p_1^{\mu_3}\bigg)\nonumber\\
	&\pi_{\mu_1}^{\lambda_1}
	\pi_{\mu_2}^{\lambda_2}  \pi_{\mu_3}^{\lambda_3}\bigg(\varepsilon^{p_1\kappa\mu_2\mu_3}p_2^{\mu_1}\bigg)=\pi_{\mu_1}^{\lambda_1}
	\pi_{\mu_2}^{\lambda_2} \pi_{\mu_3}^{\lambda_3}
	\bigg(\frac{1}{2}\left(p_1^2+p_2^2-p_3^2\right)\varepsilon^{\kappa\mu_1\mu_2\mu_3}+\varepsilon^{p_1\kappa\mu_1\mu_2}p_1^{\mu_3}+\varepsilon^{p_1\mu_1\mu_2\mu_3}p_2^\kappa\bigg)
\nonumber	\\
	&\pi_{\mu_1}^{\lambda_1}
	\pi_{\mu_2}^{\lambda_2}  \pi_{\mu_3}^{\lambda_3}\bigg(\varepsilon^{p_2\kappa\mu_1\mu_3}p_3^{\mu_2}\bigg)=\pi_{\mu_1}^{\lambda_1}
	\pi_{\mu_2}^{\lambda_2}  \pi_{\mu_3}^{\lambda_3}\bigg(\frac{1}{2}\left(p_1^2+p_2^2-p_3^2\right)\varepsilon^{\kappa\mu_1\mu_2\mu_3}+\varepsilon^{p_2\mu_1\mu_2\mu_3}p_1^\kappa-\varepsilon^{p_2\kappa\mu_1\mu_2}p_1^{\mu_3}\bigg)
\nonumber	\\
	&\pi_{\mu_1}^{\lambda_1}
	\pi_{\mu_2}^{\lambda_2}  \pi_{\mu_3}^{\lambda_3}\bigg(\varepsilon^{p_2\kappa\mu_2\mu_3}p_2^{\mu_1}\bigg)=\pi_{\mu_1}^{\lambda_1}
	\pi_{\mu_2}^{\lambda_2} \pi_{\mu_3}^{\lambda_3}
\bigg(-p_2^2\varepsilon^{\kappa\mu_1\mu_2\mu_3}+\varepsilon^{p_2\kappa\mu_1\mu_2}p_1^{\mu_3}+\varepsilon^{p_2\mu_1\mu_2\mu_3}p_2^\kappa\bigg).
\end{align}
Then, considering the identity
\begin{equation}
\varepsilon^{[\mu_2\mu_3 p_1p_2} \delta^{\mu_1]\alpha}=0
\end{equation}
and contracting with $\delta_\alpha^\kappa$, $p_{1\alpha}$ and $p_{2\alpha}$ we find
\begin{align}
	&\pi_{\mu_1}^{\lambda_1}
	\pi_{\mu_2}^{\lambda_2} \pi_{\mu_3}^{\lambda_3}
	\bigg(
	\varepsilon^{p_1p_2\mu_1\mu_3}p_3^{\mu_2}\bigg)=\pi_{\mu_1}^{\lambda_1}
	\pi_{\mu_2}^{\lambda_2} \pi_{\mu_3}^{\lambda_3}
	\bigg(-\frac{p_1^2+p_2^2-p_3^2}{2}\varepsilon^{p_1\mu_1\mu_2\mu_3}-p_1^2\varepsilon^{p_2\mu_1\mu_2\mu_3}-\varepsilon^{p_1p_2\mu_1\mu_2}p_1^{\mu_3}\bigg)
	\nonumber \\
	&\pi_{\mu_1}^{\lambda_1}
	\pi_{\mu_2}^{\lambda_2} \pi_{\mu_3}^{\lambda_3}
	\bigg(
	\varepsilon^{p_1p_2\mu_2\mu_3}p_2^{\mu_1}\bigg)=\pi_{\mu_1}^{\lambda_1}
	\pi_{\mu_2}^{\lambda_2} \pi_{\mu_3}^{\lambda_3}
	\bigg(\frac{p_1^2+p_2^2-p_3^2}{2}\varepsilon^{p_2\mu_1\mu_2\mu_3}+p_2^2\varepsilon^{p_1\mu_1\mu_2\mu_3}+\varepsilon^{p_1p_2\mu_1\mu_2}p_1^{\mu_3}\bigg)\nonumber\\
	&\pi_{\mu_1}^{\lambda_1}
	\pi_{\mu_2}^{\lambda_2} \pi_{\mu_3}^{\lambda_3}
	\bigg(\varepsilon^{p_1p_2\mu_1\mu_2}\delta^{\kappa\mu_3}\bigg)=\pi_{\mu_1}^{\lambda_1}
	\pi_{\mu_2}^{\lambda_2} \pi_{\mu_3}^{\lambda_3}
	\bigg(-\varepsilon^{p_2\mu_1\mu_2\mu_3}p_1^\kappa+\varepsilon^{p_1\mu_1\mu_2\mu_3}p_2^\kappa-\varepsilon^{p_1p_2\mu_2\mu_3}\delta^{\kappa\mu_1}+\varepsilon^{p_1p_2\mu_1\mu_3}\delta^{\kappa\mu_2}\bigg).
\end{align}
Furthermore, we need to consider the identity
\begin{equation}
\varepsilon^{[\mu_2\kappa p_1p_2} \delta^{\mu_1]\alpha}=0
\end{equation}
that once it is contracted with $p_{1\alpha}$ and $p_{2\alpha}$ we have
\begin{align}
	&
	\pi_{\mu_1}^{\lambda_1}
	\pi_{\mu_2}^{\lambda_2} \pi_{\mu_3}^{\lambda_3}
	\bigg(\varepsilon^{p_1p_2\kappa\mu_1}p_3^{\mu_2}\bigg)=\pi_{\mu_1}^{\lambda_1}
	\pi_{\mu_2}^{\lambda_2} \pi_{\mu_3}^{\lambda_3}
	\bigg(\frac{1}{2}\left(p_1^2+p_2^2-p_3^2\right)\varepsilon^{p_1\kappa\mu_1\mu_2}+p_1^2\varepsilon^{p_2\kappa\mu_1\mu_2}+\varepsilon^{p_1p_2\mu_1\mu_2}p_1^\kappa\bigg),
\nonumber	\\&\pi_{\mu_1}^{\lambda_1}
	\pi_{\mu_2}^{\lambda_2} \pi_{\mu_3}^{\lambda_3}
	\bigg(
	\varepsilon^{p_1p_2\kappa\mu_2}p_2^{\mu_1}\bigg)=\pi_{\mu_1}^{\lambda_1}
	\pi_{\mu_2}^{\lambda_2} \pi_{\mu_3}^{\lambda_3}
	\bigg(-\frac{1}{2}\left(p_1^2+p_2^2-p_3^2\right)\varepsilon^{p_2\kappa\mu_1\mu_2}-p_2^2\varepsilon^{p_1\kappa\mu_1\mu_2}+\varepsilon^{p_1p_2\mu_1\mu_2}p_2^\kappa\bigg),
\end{align}
and it is worth mentioning that the possible contraction with $\delta_\alpha^{\mu_3}$ give again an identity that is not independent taking in consideration the previous tensor identities found.

Finally, we have 
\begin{align}
\varepsilon^{[\kappa\mu_3 p_1p_2} \delta^{\mu_1]\mu_2}&=0,\\
\varepsilon^{[\kappa\mu_3 p_1p_2} \delta^{\mu_2]\mu_1}&=0,
\end{align}
giving 
\begin{align}
	&	\pi_{\mu_1}^{\lambda_1}
	\pi_{\mu_2}^{\lambda_2} \pi_{\mu_3}^{\lambda_3}
	\bigg(\varepsilon^{p_1p_2\mu_1\mu_3}\delta^{\mu_2 \kappa}\bigg)=\pi_{\mu_1}^{\lambda_1}
	\pi_{\mu_2}^{\lambda_2} \pi_{\mu_3}^{\lambda_3}
	\bigg(\varepsilon^{p_2\kappa\mu_1\mu_3}p_3^{\mu_2}+\varepsilon^{p_1p_2\kappa\mu_3}\delta^{\mu_1\mu_2}-\varepsilon^{p_1p_2\kappa\mu_1}\delta^{\mu_2\mu_3}\bigg),
\nonumber	\\
	&\pi_{\mu_1}^{\lambda_1}
	\pi_{\mu_2}^{\lambda_2} \pi_{\mu_3}^{\lambda_3}
	\bigg(\varepsilon^{p_1p_2\mu_2\mu_3}\delta^{\mu_1 \kappa}\bigg)=	\pi_{\mu_1}^{\lambda_1}
	\pi_{\mu_2}^{\lambda_2} \pi_{\mu_3}^{\lambda_3}
	\bigg(\varepsilon^{p_1\kappa\mu_2\mu_3}p_2^{\mu_1}+\varepsilon^{p_1p_2\kappa\mu_3}\delta^{\mu_1\mu_2}-\varepsilon^{p_1p_2\kappa\mu_2}\delta^{\mu_1\mu_3}\bigg),
\end{align}
and the two identities
\begin{align}
	\varepsilon^{[\kappa\mu_3 p_1p_2} \,p_2^{\mu_1]}&=0,\\
	\varepsilon^{[\kappa\mu_3 p_1p_2} \,p_1^{\mu_2]}&=0,
\end{align}
giving the only independent symmetric constraint
		\begin{align}
			\pi_{\mu_1}^{\lambda_1}
			\pi_{\mu_2}^{\lambda_2} \pi_{\mu_3}^{\lambda_3}
			\bigg(\varepsilon^{p_1p_2\kappa\mu_3}&p_2^{\mu_1}p_3^{\mu_2}\bigg)=
			\notag\\
			=\pi_{\mu_1}^{\lambda_1}
			\pi_{\mu_2}^{\lambda_2} \pi_{\mu_3}^{\lambda_3}
			\bigg\{\frac{1}{2}\Big[&-\frac{1}{2}
			\left(p_1^2+p_2^2-p_3^2\right)\varepsilon^{p_1\kappa\mu_2\mu_3}p_2^{\mu_1}-p_1^2\varepsilon^{p_2\kappa\mu_2\mu_3}p_2^{\mu_1}-\varepsilon^{p_1p_2\mu_2\mu_3}p_1^\kappa p_2^{\mu_1}
			-\varepsilon^{p_1p_2\kappa\mu_2}p_1^{\mu_3}p_2^{\mu_1}
			\notag\\
			&-\frac{1}{2}\left(p_1^2+p_2^2-p_3^2\right)\varepsilon^{p_2\kappa\mu_1\mu_3}p_3^{\mu_2}
		    -p_2^2\varepsilon^{p_1\kappa\mu_1\mu_3}p_3^{\mu_2}
			+\varepsilon^{p_1p_2\mu_1\mu_3}p_2^\kappa p_3^{\mu_2}
			-\varepsilon^{p_1p_2\kappa\mu_1}p_1^{\mu_3}p_3^{\mu_2}
			\Big]\bigg\}.
		\end{align}
	
In summary, from the analysis above, we conclude that the minimal number of tensor structures to describe $X^{\kappa\mu_1\mu_2\mu_3}$ in \eqref{kjjj} is twelve and in particular we have
\begin{align}
	&\pi_{\mu_1}^{\lambda_1}(p_1)
	\pi_{\mu_2}^{\lambda_2} (p_2) \pi_{\mu_3}^{\lambda_3}
	(p_3) \bigg(\mathcal{K}^\kappa\braket{j^{\mu_1}(p_1)j^{\mu_2}(p_2)j_5^{\mu_3}(p_3)}\bigg)=\notag\\
	&=\pi_{\mu_1}^{\lambda_1}(p_1)
	\pi_{\mu_2}^{\lambda_2} (p_2) \pi_{\mu_3}^{\lambda_3}
	(p_3) \bigg[\,p_1^\kappa\bigg(\,C_{11}\,\varepsilon^{ \mu_1\mu_2\mu_3p_1}+\,C_{12}\,
	\varepsilon^{\mu_1\mu_2\mu_3p_2}+\,C_{13}\,
	\varepsilon^{\mu_1\mu_2p_1p_2}p_1^{\mu_3}\bigg)\notag\\
	&\hspace{1cm}+p_2^\kappa\bigg(\,C_{21}\,\varepsilon^{ \mu_1\mu_2\mu_3p_1}+\,C_{22}\,
	\varepsilon^{\mu_1\mu_2\mu_3p_2}+\,C_{23}\,
	\varepsilon^{\mu_1\mu_2p_1p_2}p_1^{\mu_3}\bigg)+C_{31}\varepsilon^{\kappa\mu_1\mu_2\mu_3}+C_{32}\varepsilon^{\kappa\mu_1\mu_2 p_1}p_1^{\mu_3}\notag\\
	&\hspace{4cm}+C_{33}\varepsilon^{\kappa\mu_1\mu_2 p_2}p_1^{\mu_3}+C_{34}\varepsilon^{\kappa\mu_1p_1p_2}\delta^{\mu_2\mu_3}+C_{35}\varepsilon^{\kappa\mu_2p_1p_2}\delta^{\mu_1\mu_3}+C_{36}\varepsilon^{\kappa\mu_3p_1p_2}\delta^{\mu_1\mu_2}
	\bigg].
\end{align}
\section{Discontinuities in 3-point functions} 
The presence of a sum rule  related to the anomaly is connected with the behaviour of the spectral density for the triangle diagram and its scalar 3-point function $\mathcal{C}_0$
\beqa
\label{dis}
 \mathcal C_0(k^2, m^2)&=& \frac{1}{i \pi^2} 
 \int\frac{ d^4 l}  {(l^2 - m^2)((l-k)^2- m^2) (l-p)^2 - m^2 }
 \eeqa
with the spectral density
\beq
\rho(k^2,m^2)=\frac{1}{2 i} \textrm{Disc}\, \mathcal C_0(k^2, m^2) \,,
\label{spect}
\eeq
computed from its discontinuity
with the usual $i\epsilon$ prescription ($\epsilon >0$)
\beq
\label{discdef}
 \textrm{Disc} \, \mathcal C_0(k^2, m^2)\equiv \mathcal C_0(k^2 +i \epsilon,m^2) - \mathcal C_0(k^2 -i \epsilon,m^2).
\eeq
We can use the unitarity cutting rules to compute it
\beqa
\label{disco}
\textrm{Disc} \, \mathcal C_0(k^2, m^2)&=& \frac{1}{i \pi^2} 
 \int d^4 l \frac{2 \pi i \delta_+(l^2 - m^2) 2 \pi i \delta_+((l - k)^2- m^2)}{(l-p)^2 - m^2 + i \epsilon} \nn\\
 &=&  \frac{2 \pi}{i k^2}\log\left(\frac{1 + \sqrt{\tau(k^2 ,m^2)}}{1-\sqrt{\tau(k^2 ,m^2)}}\right)\theta(k^2 - 4 m^2) \,,
\label{cut}
 \eeqa
where $\tau(k^2,m^2) = \sqrt{1 -4m^2/k^2}$.
An alternative computation is to derive the discontinuity from the general expression of $\mathcal C_0(k^2,m^2)$ \beqa  
\label{ti}
 \mathcal C_0(k^2\pm i\epsilon,m^2) = \left\{ 
\begin{array}{ll} 
\frac{1}{2 k^2}\log^2 \frac{\sqrt{\tau(k^2,m^2)}+1}{\sqrt{\tau(k^2,m^2)}-1} & \mbox{for} \quad k^2 < 0 \,, \\
- \frac{2}{k^2} \arctan^2{\frac{1}{\sqrt{-\tau(k^2,m^2)}}} & \mbox{for} \quad 0 < k^2 < 4 m^2 \,, \\
\frac{1}{2 k^2} \left( \log \frac{1 + \sqrt{\tau(k^2,m^2)}}{1 - \sqrt{\tau(k^2,m^2)} } \mp i \, \pi\right)^2 & \mbox{for} \quad k^2 > 4 m^2 \,.  
\end{array} 
\right.
\eeqa
From the two branches encountered with the $\pm i \epsilon$ prescriptions, the discontinuity is then present only for $k^2> 4m^2$ and agrees with Eq. (\ref{disco}).
The dispersive representation of $\mathcal C_0(k^2,m^2)$ in this case is written as
\beq
\mathcal C_0(k^2,m^2)=\frac{1}{\pi} \int_{4 m^2}^{\infty} d s \frac{\rho(s, m^2)}{s - k^2 }, 
\eeq
with $\rho(s,m^2)$ given by Eqs. (\ref{spect}) and (\ref{cut}). The equation above allows to reconstruct the scalar integral 
$\mathcal C_0(k^2,m^2)$ from its dispersive part. One can prove the sum rule 
in \eqref{deltaf} by direct integration of $\rho$. The area under the integral is conserved and in the chiral limit $\rho(s,m)\to \delta(s)$. 
\section{Chern-Simons terms}
Introducing external gauge fields $B_\lambda$ and $A_\mu$, the effective action for a chiral anomaly interaction, expressed in terms of such external fields can be modifed by CS terms of the form 
\beq
V_{CS}\equiv  i \int dx A^{\lambda}(x) B^{\nu}(x) F^{A}_{\rho \sigma}(x) \varepsilon^{\lambda \nu \rho \sigma}.
\eeq
By a simple manipulation 
\beqa
V_{CS}
&=& \int dx \, dy \, dz \,  i \left( \frac{\partial}{\partial x^{\alpha}}- \frac{\partial}{\partial y^{\alpha}} \right) 
\left( \int  \frac{dk_1}{(2 \pi)^4} \frac{dk_2}{(2 \pi)^4} e^{- i p_1 (x - z) - i p_2 (y - z)} \right) B^{\lambda}(z) 
A^{\mu}(x) A^{\nu}(y) \varepsilon^{\lambda \mu \nu\alpha }     \nonumber\\
&=& (-i) \int dx \, dy \, dz \, \delta (x-z) \delta (y-z) B^{\lambda}(z)
\left(  \frac{\partial}{\partial x^{\alpha}}A^{\mu}(x)  A^{\nu}(y) - 
\frac{\partial}{\partial y^{\alpha}}A^{\nu}(y)  A^{\mu}(x)\right) \varepsilon^{\lambda \mu \nu \alpha} \nonumber\\
&=& (- i)  \int dx \, dy \, dz \,  \int \frac{dk_1 \, dp_2}{(2 \pi)^8} e^{- i p_1 (x - z) - i p_2 (y - z)} 
B^{\lambda}(z)\left(  \frac{\partial}{\partial x^{\alpha}} A^{\mu}(x)  A^{\nu}(y) - \frac{\partial}{\partial y^{\alpha}} 
A^{\nu}(y)  A^{\mu}(x) \right)  \varepsilon^{\lambda \mu \nu \alpha} \nonumber\\
&=&\int dx \, dy \, dz \, \int \frac{dp_1}{(2 \pi)^4} \frac{dp_2}{(2 \pi)^4} e^{- i p_1 (x - z) - i p_2 (y - z)} 
 \, \varepsilon^{\lambda \mu \nu \alpha} \,( k^{\alpha}_1 -k_2^\alpha)\, B^{\lambda}(z) A^{\mu}(x) A^{\nu}(y)   \nonumber\\
\eeqa
with 
\beq
\varepsilon^{\lambda \mu \nu \alpha} \,( p^{\alpha}_1 -p_2^\alpha)
\eeq
identifying the CS vertex.
If we proceed with a specific momentum parameterization of the loop, in a given parameterization, we obtain 
\beqa
p_{1\mu} \mathcal{W}^{\la\mu\nu}(p_1,p_2) = a_1 \epsilon^{\lambda\nu\alpha\beta} 
p_1^\alpha p_2^\beta \nonumber \\
p_{2\nu}\mathcal{W}^{\la\mu\nu}(p_1,p_2) = a_2 \epsilon^{\lambda\mu\alpha\beta} 
p_2^\alpha p_1^\beta \nonumber \\
p_{3\,{\la}} \mathcal{W}^{\la\mu\nu}(p_1,p_2) = a_3 \epsilon^{\mu\nu\alpha\beta} 
p_1^\alpha p_2^\beta, \nonumber \\
\eeqa
where
\beq
 a_1=-\frac{i}{8 \pi^2} \qquad a_2=-\frac{i}{8 \pi^2} \qquad a_3=-\frac{i}{4 \pi^2}.
\label{basic}
\eeq
Notice that $a_1=a_2 $, as expected from the Bose symmetry of the two {\bf V} lines. It is also well known that the total anomaly 
$a_1+a_2 + a_3 \equiv  a_n$ is regularization scheme independent. 
We recall that a shift of the momentum in the integrand $(p\rightarrow p + a)$ where $a$ is the most general momentum written in terms of the two independent external momenta of the triangle diagram $(a=\alpha (p_1 + p_2) + \beta(p_1 - p_2))$ induces on  $\Delta$ changes that 
appear only through a dependence on one of the two parameters characterizing $a$, that is 

\beq
\mathcal{W}^{\la\mu\nu}(\beta,p_1,p_2)=\mathcal{W}^{\la\mu\nu}(p_1,p_2) - \frac{i}{4 \pi^2}\beta \epsilon^{\lambda\mu\nu\sigma}\left( p_{1\sigma} - 
p_{2\sigma}\right).
\eeq

We have introduced the notation $\mathcal{W}^{\la\mu\nu} (\beta,p_1,p_2)$ to denote the shifted 3-point function, while 
$\mathcal{W}^{\la\mu\nu}(p_1,p_2)$ denotes the original one, with a 
vanishing shift.
\beqa
p_{1\mu}\mathcal{W}^{\lambda\mu\nu}(\beta',p_1,p_2)&=& (a_1 -\frac{i \beta'}{4 \pi^2})
\varepsilon^{\lambda\nu\alpha\beta}p_1^\alpha p_2^\beta,\nonumber\\
p_{2\nu}\mathcal{W}^{\lambda\mu\nu}(\beta',p_1,p_2)&=&(a_2-\frac{i \beta'}{4 \pi^2})
\varepsilon^{\lambda\mu\alpha\beta}p_2^\alpha p_1^\beta,\nonumber\\
k_\lambda\mathcal{W}^{\lambda\mu\nu}(\beta',p_1,p_2)&=&(a_3+\frac{i \beta'}{2 \pi^2})
\varepsilon^{\mu\nu\alpha\beta}p_1^\alpha p_2^\beta.
\label{bbshift}
\eeqa

\providecommand{\href}[2]{#2}\begingroup\raggedright\endgroup


\begin{thebibliography}{10}

\bibitem{Osborn:1993cr}
H.~Osborn and A.~C. Petkou, {\it {Implications of Conformal Invariance in Field
  Theories for General Dimensions}},  {\em Ann. Phys.} {\bf 231} (1994)
  311--362, [\href{http://xxx.lanl.gov/abs/hep-th/9307010}{{\tt
  hep-th/9307010}}].

\bibitem{Erdmenger:1996yc}
J.~Erdmenger and H.~Osborn, {\it {Conserved currents and the energy momentum
  tensor in conformally invariant theories for general dimensions}},  {\em
  Nucl.Phys.} {\bf B483} (1997) 431--474,
  [\href{http://xxx.lanl.gov/abs/hep-th/0103237}{{\tt hep-th/0103237}}].

\bibitem{Schreier:1971um}
E.~J. Schreier, {\it {Conformal symmetry and three-point functions}},  {\em
  Phys. Rev. D} {\bf 3} (1971) 980--988.

\bibitem{Erlich:1996mq}
J.~Erlich and D.~Z. Freedman, {\it {Conformal symmetry and the chiral
  anomaly}},  {\em Phys. Rev. D} {\bf 55} (1997) 6522--6537,
  [\href{http://xxx.lanl.gov/abs/hep-th/9611133}{{\tt hep-th/9611133}}].

\bibitem{Adler:1969er}
S.~L. Adler and W.~A. Bardeen, {\it {Absence of higher order corrections in the
  anomalous axial vector divergence equation}},  {\em Phys. Rev.} {\bf 182}
  (1969) 1517--1536.

\bibitem{Vainshtein:2002nv}
A.~Vainshtein, {\it {Perturbative and nonperturbative renormalization of
  anomalous quark triangles}},  {\em Phys. Lett.} {\bf B569} (2003) 187--193,
  [\href{http://xxx.lanl.gov/abs/hep-ph/0212231}{{\tt hep-ph/0212231}}].

\bibitem{Crewther:1972kn}
R.~J. Crewther, {\it {Nonperturbative evaluation of the anomalies in low-energy
  theorems}},  {\em Phys. Rev. Lett.} {\bf 28} (1972) 1421.

\bibitem{Broadhurst:1993ru}
D.~J. Broadhurst and A.~L. Kataev, {\it {Connections between deep inelastic and
  annihilation processes at next to next-to-leading order and beyond}},  {\em
  Phys. Lett.} {\bf B315} (1993) 179--187,
  [\href{http://xxx.lanl.gov/abs/hep-ph/9308274}{{\tt hep-ph/9308274}}].

\bibitem{Gabadadze:1995ei}
G.~T. Gabadadze and A.~L. Kataev, {\it {On connection between coefficient
  functions for deep inelastic and annihilation processes}},  {\em JETP Lett.}
  {\bf 61} (1995) 448--452, [\href{http://xxx.lanl.gov/abs/hep-ph/9502384}{{\tt
  hep-ph/9502384}}].

\bibitem{Gabadadze:2017ujx}
G.~Gabadadze and G.~Tukhashvili, {\it {Holographic CBK Relation}},  {\em Phys.
  Lett. B} {\bf 782} (2018) 202--209,
  [\href{http://xxx.lanl.gov/abs/1712.0992}{{\tt arXiv:1712.09921}}].

\bibitem{Jegerlehner:2005fs}
F.~Jegerlehner and O.~V. Tarasov, {\it {Explicit results for the anomalous
  three point function and non-renormalization theorems}},  {\em Phys. Lett.}
  {\bf B639} (2006) 299--306,
  [\href{http://xxx.lanl.gov/abs/hep-ph/0510308}{{\tt hep-ph/0510308}}].

\bibitem{Mondejar:2012sz}
J.~Mondejar and K.~Melnikov, {\it {The VVA correlator at three loops in
  perturbative QCD}},  {\em Phys. Lett. B} {\bf 718} (2013) 1364--1368,
  [\href{http://xxx.lanl.gov/abs/1210.0812}{{\tt arXiv:1210.0812}}].

\bibitem{Giannotti:2008cv}
M.~Giannotti and E.~Mottola, {\it {The Trace Anomaly and Massless Scalar
  Degrees of Freedom in Gravity}},  {\em Phys. Rev.} {\bf D79} (2009) 045014,
  [\href{http://xxx.lanl.gov/abs/0812.0351}{{\tt arXiv:0812.0351}}].

\bibitem{2009PhLB..682..322A}
R.~{Armillis}, C.~{Corian{\`o}}, and L.~{Delle Rose}, {\it {Anomaly poles as
  common signatures of chiral and conformal anomalies}},  {\em Physics Letters
  B} {\bf 682} (Dec., 2009) 322--327,
  [\href{http://xxx.lanl.gov/abs/0909.4522}{{\tt arXiv:0909.4522}}].

\bibitem{Coriano:2014gja}
C.~Corian\`o, A.~Costantini, L.~Delle~Rose, and M.~Serino, {\it {Superconformal
  sum rules and the spectral density flow of the composite dilaton (ADD)
  multiplet in $\mathcal{N}=1$ theories}},  {\em JHEP} {\bf 06} (2014) 136,
  [\href{http://xxx.lanl.gov/abs/1402.6369}{{\tt arXiv:1402.6369}}].

\bibitem{Mottola:2019nui}
E.~Mottola and A.~V. Sadofyev, {\it {Chiral Waves on the Fermi-Dirac Sea:
  Quantum Superfluidity and the Axial Anomaly}},  {\em Nucl. Phys. B} {\bf 966}
  (2021) 115385, [\href{http://xxx.lanl.gov/abs/1909.0197}{{\tt
  arXiv:1909.01974}}].

\bibitem{Ferrer:2020ulz}
E.~J. Ferrer and V.~de~la Incera, {\it {Axion-Polariton in the Magnetic Dual
  Chiral Density Wave Phase of Dense QCD}},
  \href{http://xxx.lanl.gov/abs/2010.0231}{{\tt arXiv:2010.02314}}.

\bibitem{Chernodub:2021nff}
M.~N. Chernodub, Y.~Ferreiros, A.~G. Grushin, K.~Landsteiner, and M.~A.~H.
  Vozmediano, {\it {Thermal transport, geometry, and anomalies}},  {\em Phys.
  Rept.} {\bf 977} (2022) 1--58, [\href{http://xxx.lanl.gov/abs/2110.0547}{{\tt
  arXiv:2110.05471}}].

\bibitem{Arjona:2019lxz}
V.~Arjona, M.~N. Chernodub, and M.~A.~H. Vozmediano, {\it {Fingerprints of the
  conformal anomaly on the thermoelectric transport in Dirac and Weyl
  semimetals: Result from a Kubo formula}},  {\em Phys. Rev.} {\bf B99} (2019)
  235123, [\href{http://xxx.lanl.gov/abs/1902.0235}{{\tt arXiv:1902.02358}}].

\bibitem{Bzowski:2013sza}
A.~Bzowski, P.~McFadden, and K.~Skenderis, {\it {Implications of conformal
  invariance in momentum space}},  {\em JHEP} {\bf 03} (2014) 111,
  [\href{http://xxx.lanl.gov/abs/1304.7760}{{\tt arXiv:1304.7760}}].

\bibitem{Coriano:2018bbe}
C.~Corian\`o and M.~M. Maglio, {\it {Exact Correlators from Conformal Ward
  Identities in Momentum Space and the Perturbative $TJJ$ Vertex}},  {\em Nucl.
  Phys.} {\bf B938} (2019) 440--522,
  [\href{http://xxx.lanl.gov/abs/1802.0767}{{\tt arXiv:1802.07675}}].

\bibitem{Coriano:2018bsy}
C.~Corian\`o and M.~M. Maglio, {\it {The general 3-graviton vertex ($TTT$) of
  conformal field theories in momentum space in $d =4$}},  {\em Nucl. Phys.}
  {\bf B937} (2018) 56--134, [\href{http://xxx.lanl.gov/abs/1808.1022}{{\tt
  arXiv:1808.10221}}].

\bibitem{Bzowski:2017poo}
A.~Bzowski, P.~McFadden, and K.~Skenderis,
{\it {Renormalised 3-point functions of stress tensors and conserved currents in CFT}},
{\em JHEP} {\bf11} (2018) 153,
[\href{http://xxx.lanl.gov/abs/1711.09105}{{\tt arXiv:1711.09105}}].

\bibitem{Bzowski:2018fql}
A.~Bzowski, P.~McFadden, and K.~Skenderis,
{\it {Renormalised CFT 3-point functions of scalars, currents and stress tensors}},
{\em JHEP} {\bf 11} (2018) 159,
[\href{http://xxx.lanl.gov/abs/1805.12100}{{\tt arXiv:1805.12100}}].

\bibitem{Coriano:2013jba}
C.~Coriano, L.~Delle Rose, E.~Mottola and M.~Serino,
{\it {Solving the Conformal Constraints for Scalar Operators in Momentum Space and the Evaluation of Feynman's Master Integrals}},
{\em JHEP} {\bf 07} (2013) 011,
[\href{http://xxx.lanl.gov/abs/1304.6944}{{\tt arXiv:1304.6944}}].

\bibitem{Jain:2021gwa}
S.~Jain and R.~R.~John,
{\it {Relation between parity-even and parity-odd CFT correlation functions in three dimensions}},
{\em JHEP} {\bf 12} (2021) 067,
[\href{http://xxx.lanl.gov/abs/2107.00695}{{\tt arXiv:2107.00695}}].

\bibitem{Jain:2021wyn}
S.~Jain, R.~R.~John, A.~Mehta, A.~A.~Nizami and A.~Suresh,
{\it {Momentum space parity-odd CFT 3-point functions}},
{\em JHEP} {\bf 08} (2021) 089,
[\href{http://xxx.lanl.gov/abs/2101.11635}{{\tt arXiv:2101.11635}}].

\bibitem{Marotta:2022jrp}
R.~Marotta, K.~Skenderis, and M.~Verma, {\it {Momentum space CFT correlators of
  non-conserved spinning operators}},
  \href{http://xxx.lanl.gov/abs/2212.13135}{{\tt arXiv:2212.13135}}.

\bibitem{Nishikawa:2023zcv}
K.~Nishikawa,
{\it {Conformal bootstrap in momentum space without large N}},
 \href{http://xxx.lanl.gov/abs/2303.10534}{{\tt arXiv:2303.10534}}.

\bibitem{Rosenberg:1962pp}
L.~Rosenberg, {\it {Electromagnetic interactions of neutrinos}},  {\em Phys.
  Rev.} {\bf 129} (1963) 2786--2788.

\bibitem{Knecht:2003xy}
M.~Knecht, S.~Peris, M.~Perrottet, and E.~de~Rafael, {\it {New
  nonrenormalization theorems for anomalous three point functions}},  {\em
  JHEP} {\bf 03} (2004) 035,
  [\href{http://xxx.lanl.gov/abs/hep-ph/0311100}{{\tt hep-ph/0311100}}].

\bibitem{Dolgov:1971ri}
A.~D. Dolgov and V.~I. Zakharov, {\it {On Conservation of the axial current in
  massless electrodynamics}},  {\em Nucl. Phys.} {\bf B27} (1971) 525--540.

\bibitem{Armillis:2009pq}
R.~Armillis, C.~Corian\`{o}, and L.~Delle~Rose, {\it {Conformal Anomalies and
  the Gravitational Effective Action: The $TJJ$ Correlator for a Dirac
  Fermion}},  {\em Phys. Rev.} {\bf D81} (2010) 085001,
  [\href{http://xxx.lanl.gov/abs/0910.3381}{{\tt arXiv:0910.3381}}].

\bibitem{Coriano:2021nvn}
C.~Corian\`o, M.~M. Maglio, and D.~Theofilopoulos, {\it {The conformal anomaly
  action to fourth order (4T) in $d=4$ in momentum space}},  {\em Eur. Phys. J.
  C} {\bf 81} (2021), no.~8 740, [\href{http://xxx.lanl.gov/abs/2103.13957}{{\tt
  arXiv:2103.1395}}].

\bibitem{Coriano:2022jkn}
C.~Corian\`o, M.~M. Maglio, and R.~Tommasi, {\it {Four-point functions of
  gravitons and conserved currents of CFT in momentum space: testing the
  nonlocal action with the TTJJ}},
  \href{http://xxx.lanl.gov/abs/2212.1277}{{\tt arXiv:2212.12779}}.

\bibitem{Armillis:2007tb}
R.~Armillis, C.~Corian\`o, and M.~Guzzi, {\it {Trilinear Anomalous Gauge
  Interactions from Intersecting Branes and the Neutral Currents Sector}},
  {\em JHEP} {\bf 05} (2008) 015,
  [\href{http://xxx.lanl.gov/abs/0711.3424}{{\tt arXiv:0711.3424}}].

\bibitem{Bzowski:2015yxv}
A.~Bzowski, P.~McFadden, and K.~Skenderis, {\it {Evaluation of conformal
  integrals}},  {\em JHEP} {\bf 02} (2016) 068,
  [\href{http://xxx.lanl.gov/abs/1511.0235}{{\tt arXiv:1511.02357}}].

\bibitem{Bzowski:2020lip}
A.~Bzowski, {\it {TripleK: A Mathematica package for evaluating triple-K
  integrals and conformal correlation functions}},  {\em Comput. Phys. Commun.}
  {\bf 258} (2021) 107538, [\href{http://xxx.lanl.gov/abs/2005.1084}{{\tt
  arXiv:2005.10841}}].

\bibitem{Armillis:2009sm}
R.~Armillis, C.~Corian\`o, L.~Delle~Rose, and M.~Guzzi, {\it {Anomalous U(1)
  Models in Four and Five Dimensions and their Anomaly Poles}},  {\em JHEP}
  {\bf 12} (2009) 029, [\href{http://xxx.lanl.gov/abs/0905.0865}{{\tt
  arXiv:0905.0865}}].

\end{thebibliography}

\end{document}